\documentclass[12pt]{iopart}

\expandafter\let\csname equation*\endcsname=\relax
\expandafter\let\csname endequation*\endcsname=\relax
\usepackage{graphicx}
\usepackage{mathtools}
\usepackage{hyperref}
\usepackage[normalem]{ulem}
\usepackage{amsmath}
\usepackage{amssymb}
\usepackage{amsfonts}
\usepackage{mathrsfs}
\usepackage{bm, dsfont}
\usepackage[usenames,dvipsnames]{color}
\usepackage{colortbl}
\usepackage[table]{xcolor}
\usepackage{enumitem}
\usepackage{dcolumn}
\usepackage{epstopdf}
\usepackage{cleveref}
\usepackage{hypcap}
\usepackage{verbatim, float}
\usepackage{siunitx}
\usepackage{tikz}
\usetikzlibrary{shapes.geometric, arrows}

\renewcommand{\t}[1]{\mathrm{#1}}

\newcommand{\id}{\mathds{1}}
\newcommand{\prjct}[1]{\mathinner{|{#1}\rangle}\!\!\mathinner{\langle{#1}|}}

\usepackage{xcolor}
\definecolor{navyblue}{rgb}{0.0, 0.0, 0.5}
\definecolor{orangered4}{rgb}{0.55, 0.15, 0.0}

\usepackage{listings}
\usepackage{algorithm,algorithmic}
\usepackage[algo2e,ruled,vlined]{algorithm2e}
\usepackage{caption}

\def\ket#1{\left|#1\right\rangle}
\def\bra#1{\left\langle#1\right|}
\def\scal#1#2{\langle#1|#2\rangle}

\def\abs#1{\left\lvert#1\right\rvert}

\def\s_der#1{{\rm d}\! #1}
\def\der#1{{\rm d} #1}
\def\abs#1{\left|#1\right|}
\newcommand{\norm}[1]{\left\vert \left\vert #1 \right\vert\right\vert}

\usepackage{makecell}

\newcommand{\ave}[1]{\left< #1 \right>}

\newcommand{\ii}{\mathrm{i}}
\newcommand{\DP}{$\partial$P}
\renewcommand{\vec}{\bm} 

\usepackage{booktabs}

\begin{document}
\title[]{Control of Stochastic Quantum Dynamics by Differentiable Programming}

\author{Frank Sch\"afer\textsuperscript{1}, Pavel Sekatski\textsuperscript{1,2}, Martin Koppenh\"ofer\textsuperscript{1,3}, \\ Christoph Bruder\textsuperscript{1}, and Michal Kloc\textsuperscript{1}}

\address{$^1$ Department of Physics, University of Basel, Klingelbergstrasse 82, CH-4056 Basel, Switzerland}
\address{$^2$ Department of Applied Physics, University of Geneva, CH-1211 Geneva, Switzerland}
\address{$^3$ Pritzker School of Molecular Engineering, University of Chicago, Chicago, Illinois 60637, USA} 
\ead{frank.schaefer@unibas.ch, michal.kloc@unibas.ch}
\vspace{10pt}
\begin{indented}
\item[]\today
\end{indented}

\begin{abstract}
Control of the stochastic dynamics of a quantum system is indispensable in fields such as quantum information processing and metrology.
However, there is no general ready-made approach to the design of efficient control strategies. 
Here, we propose a framework for the automated design of control schemes based on differentiable programming (\DP{}). 
We apply this approach to the state preparation and stabilization of a qubit subjected to homodyne detection. 
To this end, we formulate the control task as an optimization problem where the loss function quantifies the distance from the target state, and we employ neural networks (NNs) as controllers. 
The system's time evolution is governed by a stochastic differential equation (SDE). 
To implement efficient training, we backpropagate the gradient information from the loss function through the SDE solver using adjoint sensitivity methods. 
As a first example, we feed the quantum state to the controller and focus on different methods of obtaining gradients. 
As a second example, we directly feed the homodyne detection signal to the controller. The instantaneous value of the homodyne current contains only very limited information on the actual state of the system, masked by unavoidable photon-number fluctuations. 
Despite the resulting poor signal-to-noise ratio, we can train our controller to prepare and stabilize the qubit to a target state with a mean fidelity of around 85\%. 
We also compare the solutions found by the NN to a hand-crafted control strategy.
\end{abstract}

\section{Introduction}
\label{Sec:Introduction}

The ability to precisely prepare and manipulate quantum degrees of freedom is a prerequisite of most applications of quantum mechanics in sensing, computation, simulation and general information processing. Many relevant tasks in this area can be formulated as optimal control problems, and therefore, \textit{quantum control} is a rich and very active research field, see~\cite{dalessandro2008,wiseman09} for two recent textbooks that provide theoretical background and~\cite{glaser2015training} for a recent review of important issues.

A typical goal of quantum control is to find a sequence of operations or parameter values (e.g. external field amplitudes) such that the quantum system under consideration is maintained in a certain target state or evolves in a desired fashion subject to additional boundary conditions (e.g. most rapid evolution for a prescribed maximal strength of the control fields). 
In the case of \textit{feedback} control the control sequence is determined based on a signal coming from the system~\cite{glaser2015training,nori2015}. 
The control task and its boundary conditions are typically specified by a loss function. 
To cast the optimal control problem into an optimization task, one then introduces a parametric ansatz for feedback schemes, also known as controllers, and explores the parameter space to minimize the loss function.

Reinforcement learning (RL)~\cite{sutton1998,lillicrap2015} has been proposed as a suitable framework to develop control strategies.
In this framework, the controller (or \textit{agent}), optimizes its strategy (the \textit{policy}) based on the loss function (the \textit{rewards}) obtained from repetitive interactions with the system to be controlled.
In the context of quantum physics, RL has proven useful, e.g. for reducing the error rate in quantum gate synthesis~\cite{niu2019}, for autonomous preparation of Floquet-engineered states~\cite{bukov18PRB}, and for optimal manipulation of many-qubit systems~\cite{bukov18PRX}. 
RL is a black-box setting, i.e.\ the agent has no prior knowledge on the structure of the system it interacts with, and has to develop its policy (explore the space of control parameters) only relying on its past interactions with the system. 
This makes RL very versatile but requires a large number of training episodes to find a good strategy.

Optimal control design is rarely done on an (unknown) system in situ. 
Instead one trains the controller on a physical model of the real system. 
This means that rather than learning how to interact with an unknown  environment, one actually starts with  a lot of prior scientific knowledge about the system, namely its precise dynamical model. 
In the simplest cases, one can even solve the optimal control problem for this model analytically.
Generally, using prior information leads to more data-efficient training~\cite{schaefer20,rackauckas2020universal}.
In the context of the present paper, we use the physical model of the system to efficiently compute the loss function's gradients with respect to the parameters of the controller ansatz.
Naturally, having access to the gradients of the loss function can streamline the optimal control design tremendously.

In the case of quantum control the model consist of a quantum state space and a parametric equation of motion. The dynamics of the system can be solved for fixed values of the parameters of the model and of the controller.
Usually, this is done numerically.
Then, the most naive way to obtain the loss function's gradients is to solve the dynamics for a set of parameter values in the neighborhood of each point and use finite difference quotients. 
Yet, this method is unfeasible if the number of parameters is large, it suffers from floating-point errors, and it may be numerically unstable. 
To circumvent these issues, automatic differentiation (AD) has been proposed as another approach to calculate gradients numerically~\cite{leung17,abdelhafez19}, a paradigm also known as differentiable programming (\DP{})~\cite{rackauckas2020generalized,liao2019}. 
By backpropagating the loss function's  sensitivity through the numerical simulation, one can compute the gradients with similar time complexity as solving the system's dynamics~\cite{rackauckas2019diffeqflux}.
Recently, these techniques have been merged with deep neural networks (NNs) as ansatz for the controllers~\cite{schaefer20, Wu19, coopmans2020}. 
This is possible because the training of deep NNs is based on stochastic gradient descent through a large parameter space and becomes efficient when used in conjunction with \DP{}-compatible solvers for the system's dynamics.

In this work, we develop such a  physics-informed reinforcement learning framework based on \DP{} and NNs  to control the stochastic dynamics of a quantum system under continuous monitoring~\cite{wiseman09,breuer02}.
Continuous measurements, such as photon counting and homodyne detection, allow one to gain information on the random evolution of a dissipative quantum system~\cite{wiseman09}. 
This information can be used to estimate the state of the quantum system \cite{Briant2003,Iwasawa2013,Wieczorek2015}, to implement feedback protocols \cite{Wiseman1993,Mancini1998,Hofmann1998,Doherty1999,Wilson2015}, to generate nonclassical states \cite{Nha2004,Viviescas2010,Koppenhoefer2018,Koppenhoefer2020}, and to implement teleportation protocols \cite{Bose1999,Greplova2016}.
Continuous homodyne detection can be realized experimentally in the microwave~\cite{ficheux2018, vijay2012} and optical regime~\cite{Briant2003,Armen2002}.
The time evolution of a monitored quantum system is described by quantum trajectories, which are solutions of differential equations driven by a L\'evy process.

To illustrate our framework, we focus on a qubit subjected to continuous homodyne detection~\cite{ficheux2018,naghiloo2016} described by a stochastic Schr\"odinger equation.
We engineer a controller which  provides a control scheme to fulfil a given state preparation task based on the measured homodyne current.
This situation extends an earlier study~\cite{schaefer20}, where it has been demonstrated that \DP{} can be efficiently used for quantum control in the context of (unitary) closed-system dynamics, i.e.\ when the dynamics follows an ordinary differential equation.
The stochastic nature of the problem analyzed here renders the control task more challenging because the controller must adapt to the random evolution of the quantum state in each trajectory.
Moreover, the instantaneous value of the homodyne current does not determine the actual state of the qubit. 
It is correlated only to the projection of the state onto the $x$-axis and this signal is hidden in the noise which dominates the measured homodyne current~\cite{bouten2007,flurin2020}. 
 Thus, the information about the state of the qubit at a given time must be filtered out from the time series of measurement results.

This paper is organized as follows:
In Section~\ref{Sec:Theoretical}, we describe the proposed setup of a qubit in a leaky cavity subjected to homodyne detection and derive the stochastic Schr\"odinger equation that describes its dynamics. 
We then discuss two ways to use the record of the homodyne detection signal in a feedback scheme to engineer a drive that can be applied to the qubit to perform a desired control task, e.g. state preparation or stabilization. 
We also introduce the concept of adjoint sensitivity methods that we use to efficiently compute gradients with respect to solutions of SDEs.
Section~\ref{Sec:SDE_control_state} describes the first feedback scheme in detail: Here, we assume that the controller has direct access to the quantum state, e.g. through an appropriate filtering procedure applied to the measurement records. 
We compare three strategies, viz.\ a hand-crafted control scheme, one in which a neural network continuously updates the control drive based on the knowledge of the state, and a numerically less demanding one with a piecewise-constant control drive.
Section~\ref{Sec:SDE_control_Jhom} presents the second feedback scheme where directly the measurement record of the homodyne current is fed to the NN representing the controller. 
In this setup, the NN must first learn how to filter the data to obtain information on the state of the system.
Only after that it can propose an efficient control strategy.
We conclude in Section~\ref{Sec:Conclusion} and discuss potential future applications.

\section{Theoretical background}
\label{Sec:Theoretical}

\subsection{A qubit under homodyne detection}
\label{Subsec:Homodyne}

\begin{figure}[h!]
	\centering
	\includegraphics[width=1\linewidth, angle=0]{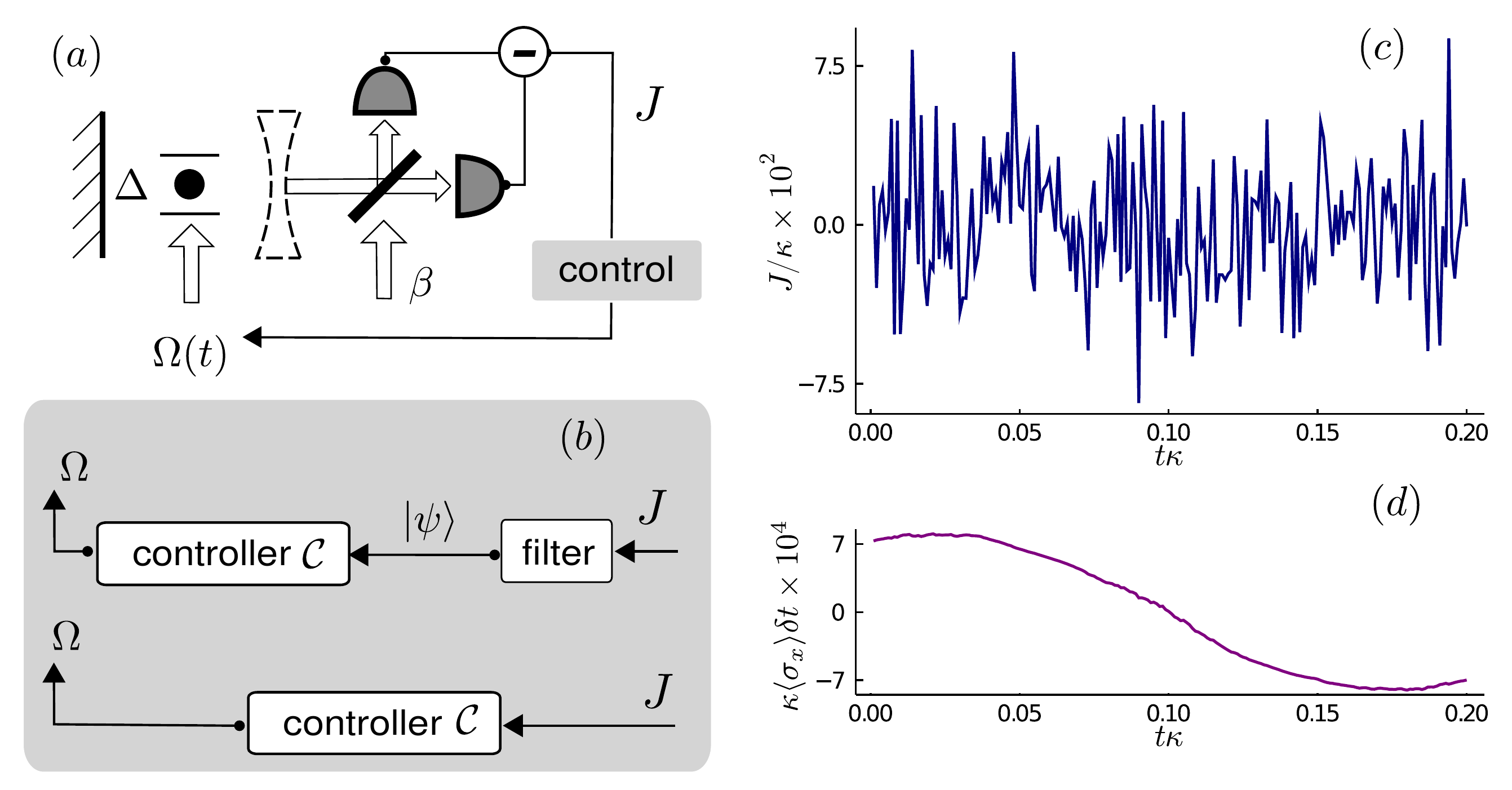}
	\caption{
	(a) Sketch of the considered setup. A qubit coupled to a leaky cavity is continuously monitored by a homodyne detection  measurement.
	The radiation emitted by the qubit is mixed with a local oscillator laser with complex amplitude $\beta$ at a beamsplitter. The signal $J(t)$, which is the difference of the photocurrents of the two detectors, is fed into a control agent, which applies a drive $\Omega(t)$ to the qubit to generate and stabilize a target state.
	(b) Different structures of the controller considered in Secs.~\ref{Sec:SDE_control_state} and~\ref{Sec:SDE_control_Jhom}, respectively. (c) Example of a typical signal $J(t)$ of the homodyne detection measurement. 
	(d) The homodyne detection signal is proportional to the quadrature $\ave{\sigma_x}$, which is hidden in the noise of the measurement [note the different axis scaling in (c) and (d)]. The data has been integrated over a measurement interval $\delta t =10^{-3} \kappa^{-1}$. The parameters of the physical model are $\Delta=20 \kappa$, $\Omega=2 \kappa$.}
	\label{Fig:1}
\end{figure}

We consider a driven two-level system with states $\ket{g}$ and $\ket{e}$.
In a rotating frame, its Hamiltonian reads 
\begin{equation}
H= \frac{\Delta}{2} \sigma_z + \frac{\Omega(t)}{2}\sigma_x \,,
\label{Eq:H}
\end{equation}
where $\sigma_x, \ \sigma_z$ are Pauli matrices, $\Omega(t)$ is the Rabi frequency of the drive laser, and $\Delta= \omega_{eg}- \omega_\textrm{laser}$ is the detuning between the qubit and the laser. 
The qubit can spontaneously decay into a photon field $a(t)$ via the interaction Hamiltonian
\begin{equation}
    H_\mathrm{int} = \ii  \sqrt{\kappa}\left[\sigma_+ a(t)  - \sigma_- a^\dag(t) \right]\,,
\end{equation}
where $\kappa$ is the decay rate.
The field operators $a(t)$ and $a^\dag(t)$ satisfy the commutation relation $[a(t),a^\dag(t')]=\delta(t-t')$. 
 We assume that the field is initially in the vacuum state,  $\langle a^\dag(t) a(t')\rangle =0$.
Physical examples of such a system are a two-level atom inside a leaky single-mode cavity that can be adiabatically eliminated, or an artificial atom, e.g.\ a superconducting qubit, coupled to a waveguide.
The  radiation emitted from the two-level system is monitored with a continuous homodyne measurement, as depicted in Fig.~\ref{Fig:1}(a).

We show in~\ref{App:Homodyne} (see also \cite{wiseman09,breuer02}) that the evolution of the qubit is governed by a stochastic Schr\"odinger equation
\begin{equation}
\der{\tilde{\ket{\psi}}}= \der{t} \left\{-\ii H-\frac{\kappa}{2}\sigma_+\sigma_- + J(t) \sigma_-\right\}\tilde{\ket{\psi}}\,,
\label{Eq:Homodyne_NoNorm}
\end{equation}
where $\tilde{\ket{\psi}}$ denotes an unnormalized qubit state.
The instantaneous value of the measured homodyne current $J(t)$ is a random variable satisfying
\begin{equation}
 J(t)=\kappa \ave{\sigma_x}_{\psi(t)}+ \sqrt{\kappa}\, \xi(t)\,.
\label{Eq:Jhom}
\end{equation} 
Here, $\ave{\sigma_x}_{\psi(t)}$ is the expectation value of $\sigma_x$ at time $t$, and $\xi(t)$ is a stochastic white-noise term satisfying $\mathds{E}[\xi(t)\xi(t')]\propto \delta(t-t')$, which stems from the shot noise of the local oscillator. 
Heuristically, $\xi(t)$ can be considered as the derivative of a stochastic Wiener increment, $\xi(t)= \der{W}(t)/\der{t}$, such that the contribution of the noise to the current integrated over a short time interval $\der{t}$ is described by a Wiener process, 
\begin{equation}
    J(t)\der{t} = \kappa \ave{\sigma_x}_{\psi(t)}\der{t} + \sqrt{\kappa}\, \der{W}(t)~.
    \label{Eq:Jhom2}
\end{equation}
The ensemble averages of the stochastic Wiener increment $\der{W}$ satisfy $\mathds{E}[\der{W}(t)] = 0$ and $\mathds{E}[\der{W}(t)^2] = \der{t}$.
Several remarks are in order to better understand the dynamics of the system.

First, Eq.~\eqref{Eq:Jhom2} implies that the value of the current over a short interval $J(t) \der{t}$  contains only very little information about the state $\ket{\psi(t)}$ of the qubit.
This can be seen from the (heuristically stated) vanishing signal-to-noise ratio 
\begin{equation}
    \frac{\kappa \ave{\sigma_x}_{\psi(t)}\der{t}}{\sqrt{\mathds{E}[\kappa \, \der{W}(t)^2]}} 
    = \frac{\kappa \ave{\sigma_x}_{\psi(t)}\der{t}}{\sqrt{\kappa \der{t}}} = \ave{\sigma_x}_{\psi(t)} \sqrt{ \kappa \der{t}}.
\end{equation}
If $\ave{\sigma_x}_{\psi(t)}$ was a constant signal, one could simply integrate the current $J(t)$ over a time interval longer than $1/\kappa$ to increase the signal-to-noise ratio.
However, this is not possible because relaxation will change the state of the two-level system on the time scale $1/\kappa$.
Thus, the low signal-to-noise ratio is an intrinsic feature of this homodyne detection scheme.
This is illustrated in Figs.~\ref{Fig:1}(c) and (d), where we show a simulation of the homodyne current $J(t)$ together with the respective value $\ave{\sigma_x}_{\psi(t)}$ for a single quantum trajectory.

Second, Eq.~\eqref{Eq:Homodyne_NoNorm} shows that the infinitesimal time evolution and thus the quantum trajectory is fully determined by the record of the measured homodyne current $\bm J_t$ and the values of the applied drive $\bm \Omega_t$, which are vectors containing the respective values of $J(t)$ and $\Omega(t)$ from the start time $t_0 = 0$ until the time $t$.
In \ref{app: filter}, we derive a closed-form expression of the operator  $D_t=D_t[\bm J_t, \bm \Omega_t]$, which gives the mapping
\begin{equation}\label{eq: filter main}
 \rho_t =\frac{D_t  \rho_0 D_t^\dag}{\tr 
[D_t  \rho_0 D_t^\dag]},
\end{equation}
between the states of the qubit at times $t_0=0$ and $t$.
The operator $D_t[\bm J_t, \bm \Omega_t]$ can be interpreted as a filter determining the state of the qubit at time $t$ from the values of the measured homodyne current and the applied drive.

Finally, Eq.~\eqref{Eq:Homodyne_NoNorm} does not preserve the norm of the state $\ket{\psi}$, as indicated by the tilde. 
This is no problem if one integrates Eq.~\eqref{Eq:Homodyne_NoNorm} numerically, since one can renormalize the state after each time step.
For analytical calculations, it is useful to add some correction terms (see \ref{App:HD:DerivationNormPreservingSDE} or \cite{wiseman09,breuer02}) such that the norm of the state is preserved up to second order in $\der{t}$,

\begin{equation}
\s_der{\ket{\psi}}= K_{\psi(t)}\der{t}+ M_{\psi(t)}\der{W}(t) \,.
\label{Eq:SDE_homodyne}
\end{equation}
Here, the nonlinear drift and diffusion terms are
\begin{align}
    K_{\psi(t)} &=\left(-\ii H +\frac{ \kappa}{2} \left\{\ave{\sigma_x}_{\psi(t)} \sigma_- - \sigma_+ \sigma_- -\frac{1}{4}\ave{\sigma_x}_{\psi(t)}^2\right\}\right) \ket{\psi}\,,
\label{Eq:Deterministic} \\
    M_{\psi(t)} &= \sqrt{\kappa} \left\{\sigma_--\frac{1}{2} \ave{\sigma_x}_{\psi(t)}\right\}\ket{\psi}\,.
\label{Eq:StochPart}
\end{align}
Note that Eq.~\eqref{Eq:SDE_homodyne} is a stochastic Schr\"odinger equation in the It\^{o} form with multiplicative scalar noise.

\subsection{Feedback control overview}
\label{Subsec: control}

In our control protocol, the results of the homodyne detection measurement determine the drive $\Omega(t)$ to be applied to the qubit. 
We will consider two different schemes which are sketched in Fig.~\ref{Fig:1}(b).

In the first scheme, discussed in Section~\ref{Sec:SDE_control_state}, we filter the homodyne signal  to extract the system's exact state $\ket{\psi(t)}$ at time $t$. 
Subsequently, the controller receives this state as an input and determines the drive $\Omega(t)$ to be applied next.
Equations~\eqref{eq: filter main} and \eqref{eq: filter app} give an explicit filtering procedure $D_t[\bm J_t, \bm \Omega_t]$ to determine the state of the qubit at time $t$ from the records of homodyne measurements $\bm J_t$ and the drive $\bm \Omega_t$, which are known in the experiment.
Since we train the agent on simulated trajectories, we know the system's state at each time step. 
Therefore, we can skip the filter in Fig.~\ref{Fig:1}(b) and directly feed back the solution of the SDE solver at each time step to our controller. 
In other words we assume a perfect filtering at any time.
Thus, this situation corresponds to the feedback  used in Ref.~\cite{schaefer20} but the deterministic evolution is replaced by an SDE.
We will use this control scheme in Section~\ref{Sec:SDE_control_state} to test different  backpropagation methods and to compare different control strategies.

In the second scheme, discussed in Section~\ref{Sec:SDE_control_Jhom}, the controller obtains at time $t$ the homodyne current record $\bm J_\tau(t)$ measured over some time interval $[t-\tau,t]$.
Now, the NN forming the controller must simultaneously learn how to filter the signal from the noise and to predict the next action $\Omega(t)$.
Such an implementation of the control protocol based only on $J(t)$ is a challenging task because the signal of the system quadrature $\ave{\sigma_x}_{\psi(t)}$ is hidden in the noise as  discussed in Section~\ref{Subsec:Homodyne}.

\subsection{Workflow}
\label{Subsec:Differentiable}

\begin{figure}[h!]
	\centering
	\includegraphics[width=1.0\linewidth, angle=0]{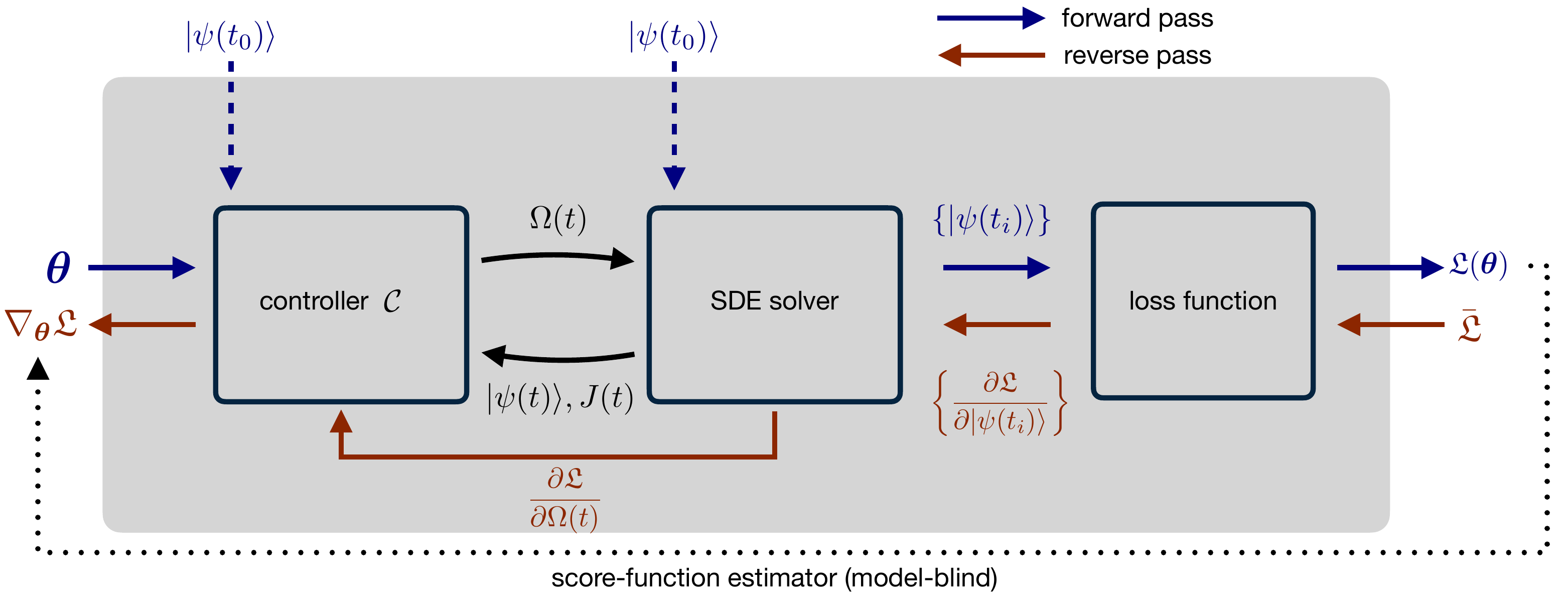}
	\caption{
	Workflow of the learning scheme discussed in Section~\ref{Subsec:Differentiable} to train the controller.
	In the forward pass, a controller, which is in this work implemented by a neural network, maps the 
	present quantum state $\ket{\psi(t)}$ (see Section~\ref{Sec:SDE_control_state}) or a measurement of the homodyne current $J(t)$ 
	(see Section~\ref{Sec:SDE_control_Jhom}) to a  drive $\Omega(t)$. Then, an SDE is solved to determine the subsequent state and homodyne detection current $J(t)$.
	A loss function $\mathfrak{L}$ modeling the state preparation objective and possible constraints is evaluated based on a quantum trajectory, i.e.\ a sequence of states. 
    In the backward pass, the gradient of the loss function with respect to the parameters $\vec{\theta}$ of the controller is evaluated by (adjoint) sensitivity methods (see Section~\ref{Subsec:Adjoint}).
	This step incorporates physical knowledge of the system into the training process and is numerically more efficient than a model-blind gradient estimation.
	The gradient of the loss function with respect to the parameters of the controller is used to update the control strategy in a series of training epochs.
	}
	\label{Fig:2}
\end{figure}

The learning scheme to control the stochastic dynamics of the continuously monitored qubit based on \DP{} consists of three building blocks as sketched in Fig.~\ref{Fig:2}: a parametrized controller $\mathcal{C}$, which will be formed by a NN, a model of the dynamics, expressed as an SDE, and a loss function.

At the beginning of each run, we initialize the system in an arbitrary state on the Bloch sphere, 
\begin{equation}
    \ket{\psi(t_0)}= \cos(\frac{{\vartheta}}{2}) \ket{e}+ \sin( \frac{{\vartheta}}{2}) \mathrm{e}^{i\phi}\ket{g},
    \label{Eq:initial_state}
\end{equation}
where $\ket{e}$ and $\ket{g}$ are the excited and ground states of the qubit in the $z$-basis, respectively.
To ensure that the controller will perform optimally for any initial state, we sample the angles $\vartheta$ and $\phi$ uniformly from their intervals $[0,\pi]$ and $[0,2\pi)$, respectively.

Depending on the chosen control scheme, cf.\ Fig.~\ref{Fig:1}(b), the controller receives as an input either the quantum state $\ket{\psi(t)}$ (Section~\ref{Sec:SDE_control_state}) or the last homodyne detection record in form of a vector $\vec{J}_\tau(t)$ gathered over some time interval $[t-\tau,t]$  (Section~\ref{Sec:SDE_control_Jhom}). 
Moreover, in Section~\ref{Sec:SDE_control_Jhom} the controller also receives a vector $\vec{\Omega}_m(t)$ of the $m$ last control actions applied prior to the time $t$.

The controller then maps this input to the next value of the drive, $\Omega(t)$.
Given $\ket{\psi(t)}$ and $\Omega(t)$, we use the Runge-Kutta Milstein solver from the StochasticDiffEq.jl package~\cite{rackauckas2017differentialequations, rackauckas2017adaptive,rackauckas_stability-optimized_2018} to calculate the next state $\ket{\psi(t+\mathrm{d}t)}$ according to the SDE~\eqref{Eq:SDE_homodyne}.
This loop between control agent and SDE solver is iterated for all time steps. 
We store the quantum states $\ket{\psi(t_i)}$ and the drive values $\Omega(t_i)$ at $N$ uniformly spaced time steps $\{t_i\}_{i=1}^N$ to be able to evaluate the loss function.
Hereafter, the set $\{\ket{\psi(t_i)}, \Omega(t_i)\}$ is called checkpoints.

We minimize a loss function of the form~\cite{schaefer20,leung17,abdelhafez19,caneva2011}
\begin{equation}
  \mathfrak{L}=\sum_{\mu} c_\mu \mathfrak{L}_\mu \,,
\label{Eq:Loss}
\end{equation} 
where the terms $\mathfrak{L}_\mu$ encode case-specific objectives of the optimization process, see, e.g.\ Ref.~\cite{leung17}.
Their relative importance can be controlled by the weights $c_\mu$.
We choose the weights $c_\mu$ empirically but, if necessary, they could also be tuned by means of hyperparameter optimization techniques~\cite{schaefer20}.  
To enforce a large fidelity with respect to the target state over the whole control interval, we include into the loss function 
the average infidelity of the checkpoints $\ket{\psi(t_i)}$ with respect to the target state $\ket{\psi_{\rm tar}}$,
\begin{equation}
\mathfrak{L}_{F}=\frac{1}{N}\sum_{i=0}^{N}{\left(1-\abs{\scal{\psi(t_i)}{\psi_{\rm tar}}}^2\right)}\,.
\label{Eq:Infidelity}
\end{equation} 
This form of the loss function also leads to a time-optimal performance of the controller.
To focus on specific time intervals, like the last few steps for example, the sum in Eq.~\eqref{Eq:Infidelity} can be straightforwardly adjusted.
We will consider the target state $\ket{\psi_{\rm tar}} = \ket{e}$ in the following.
In addition to $\mathfrak{L}_F$, we include the term
\begin{equation}
    \mathfrak{L}_{\Omega}=\frac{1}{N}\sum_{i=0}^N |\Omega(t_i)|^2
\end{equation}
in the loss function to favor smaller amplitudes of the  drive $\Omega(t_i)$. 
In Section~\ref{Sec:SDE_control_Jhom}, we will find that this term is important to suppress the collapse of the NN towards a strategy where constant maximal pulses are applied during the training.

At this stage, we have calculated a quantum trajectory and evaluated its value of the loss function. In the next step, we must update the control strategy to decrease the value of the loss function. The derivative of the loss function $\mathfrak{L}$ with respect to the parameters of the neural network provides a meaningful update rule towards a better control strategy. Thus, we need to calculate the gradient $\nabla_{\vec{\theta}} \mathfrak{L}$. This can be computed efficiently using sensitivity methods for SDEs discussed next. 

\subsection{Adjoint sensitivity methods}
\label{Subsec:Adjoint}

The loss function $\mathfrak{L} = \mathfrak{L}(\{\ket{\psi(t_i)}, \Omega(t_i)\})$, defined in Eq.~\eqref{Eq:Loss}, is a scalar function which explicitly depends on the checkpoints and implicitly on the parameters $\vec{\theta}$ of the controller.
In contrast to score-function estimators~\cite{glynn1990likelihood,yang1991monte, kleijnen1996optimization}, such as the REINFORCE algorithm~\cite{williams1992simple}, we will incorporate the physical model into the gradient computation as sketched in Fig.~\ref{Fig:2}.

Automatic differentiation (AD) is a powerful tool to evaluate gradients of numeric functions at machine precision~\cite{baydin2017automatic}. 
A key concept to AD is the computational graph~\cite{wengert1964simple}, also known as Wengert trace, which is a directed acyclic graph that represents the sequence of elementary operations that a computer program applies to its input values to calculate its output values.
The nodes of the computational graph
 are the elementary computational steps of the program called \textit{primitives}. 
The outcome of each primitive operation,  called an \textit{intermediate variable}, is evaluated in the \textit{forward pass} through the graph. 

In forward-mode AD, one associates with each intermediate variable $v_j$ the value of the derivative
\begin{align}
 \dot{v}_{ji} =\frac{\partial v_j}{\partial \theta_i}
\end{align}
with respect to a parameter $\theta_i$ of interest.
The derivatives $\dot{v}_{ji}$ are calculated together with the associated intermediate values $v_j$ in the forward pass, i.e.\ the gradient is pushed forward through the graph.
This procedure must be repeated for each parameter $\theta_i$, therefore, forward-mode AD scales poorly in computation time with increasing number of parameters $\{\theta_i\}$. 

In contrast, reverse-mode AD traverses the computation graph backwards from the loss function to the parameters $\theta_i$ by defining an \textit{adjoint process}:
\begin{equation}
\bar{v}_i = \frac{\partial \mathfrak{L}}{\partial v_i} \, ,
\end{equation}
which is the \textit{sensitivity} of the loss function $\mathfrak{L}$ with respect to changes in the intermediate variable $v_i$.
Reverse-mode AD is very efficient in terms of the number of input parameters because one needs just a single \textit{backward pass}  after the forward pass to obtain the gradient with respect to all parameters $\{\theta_i\}$. 
Thus, we always implement the AD of the NN in reverse mode.
However, reverse-mode AD might be very memory expensive because all intermediate variables $v_i$ from the forward pass need to be stored for the backward pass. 
Therefore, reverse-mode AD scales poorly in memory if the number of steps and parameters of the SDE solver increases. Whether forward-mode or reverse-mode AD is more efficient to calculate gradients of the loss function depends on the specific details of the control loop shown in Fig.~\ref{Fig:2}. 

In fact, it is possible to combine different AD methods on different parts of the computational graph as we illustrate now.
In Secs.~\ref{Subsec:Piecewise} and \ref{Sec:SDE_control_Jhom}, we will use Algorithm~\ref{algo:piecewise}, where the drive $\Omega(t)$ is piecewise constant, i.e.\ a constant value of $\Omega$ is applied over $N_{\rm sub}$ successive time steps between two checkpoints.
In this case, the memory consumption of the reverse-mode AD of the NN is moderate since the number of parameters  $\{\Omega(t_i)\}$ grows only with the number of checkpoints rather than with the number of time steps. 
In contrast, in the SDE solver the evaluation of the time evolution between two checkpoints $\{\ket{\psi(t_i)}, \Omega(t_i)\}$ and $\{\ket{\psi(t_{i+1})}, \Omega(t_{i+1})\}$ only depends on the single parameter $\Omega(t_i)$, which makes forward-mode AD very efficient~\cite{rackauckas2018comparison}. 
Therefore, we nest both methods and use an inner forward-mode AD through the SDE solver and an outer reverse-mode AD for the remaining parts, i.e. the NN and the computation of the loss function.

\begin{minipage}[h!]{0.94\linewidth}
	\begin{minipage}[t]{0.49\linewidth}
       \begin{algorithm}[H]
       \DontPrintSemicolon
       \SetAlgoLined
       \KwIn{$\psi(t_0)$, $t_0$, $\Omega(t_0)=0$}
       \KwResult{$\mathfrak{L}$}
        \For{$i=0:N-1$}{
           \emph{compute and store checkpoints:}\\
            $\Omega(t_{i+1}) \leftarrow \text{NN}_{\vec{\theta}}(\psi(t_i))$\\
           $\psi(t_{i+1}) \leftarrow $ solve($\psi(t_i)$,{$\Omega(t_{i+1})$}) for SDE \eqref{Eq:SDE_homodyne} in $[t_i, t_{i+1}]$
        }
        $\mathfrak{L}  \leftarrow$ loss($\{\psi(t_i), \Omega(t_i)\}$)
       \caption{Piecewise constant}
       \label{algo:piecewise}
       \end{algorithm}
	\end{minipage}
	\hfill
	\begin{minipage}[t]{0.49\linewidth}
	  \begin{algorithm}[H]
        \DontPrintSemicolon
        \SetAlgoLined
        \KwIn{$\psi(t_0)$, $t_0$, $\Omega(t_0)=0$}
        \KwResult{$\mathfrak{L}$}
  
        \emph{compute and store checkpoints:}\\
        $\{\psi(t_i), \Omega(t_i)\} \leftarrow$ solve($\psi(t_0)$,{$\text{NN}_{\vec{\theta}}$}) for SDE \eqref{Eq:SDE_homodyne} in $[t_0, t_{N}]$
        \vspace{15.5mm} \\
       $\mathfrak{L}  \leftarrow $ loss($\{\psi(t_i), \Omega(t_i)\}$)
       \caption{Continuously updated}
       \label{algo:continuous}
       \end{algorithm}
	\end{minipage}
\end{minipage} \vspace{0.5cm} 

Restricting oneself to a piecewise-constant control drive, however, prevents the controller from reacting instantaneously to changes of the state.
To implement a fast control loop, the controller  has to be placed in the drift term of the SDE~\eqref{Eq:SDE_homodyne}.
Thus, the parameters entering the solver will be the NN parameters $\vec{\theta}$, see Algorithm~\ref{algo:continuous}.
The \textit{continuous adjoint sensitivity method}~\cite{pontryagin2018,li2020,kidger2021} circumvents the resulting memory issues by introducing a new primitive for the whole SDE in the backward pass of the code.
This new primitive is determined by the solution of another SDE problem, the \textit{adjoint SDE}.
Formally, defining the \textit{adjoint process} as
  \begin{equation}
    \vec{a}_{\psi}(t)=\nabla_{\psi(t)} \mathfrak{L}(\{\ket{\psi(t_i)}\}) \,,
  \end{equation}
the adjoint SDE problem in the It\^o sense satisfies the differential equation
\begin{align}
     \der{\vec{a}_{\psi}}(t)=& -\left(\vec{a}_{\psi}^\dagger(t) \cdot \nabla_{\psi(t)}\right) \left(K_{\psi(t)} - 2C^{\rm IS}_{\psi(t)}\right) \der{t} 
     -\left(\vec{a}_{\psi}^\dagger(t) \cdot \nabla_{\psi(t)}\right) M_{\psi(t)}  \der{W(t)}\,,
  \label{Eq:adjoint_SDE}
\end{align}
with the initial condition 
 \begin{align}
    \vec{a}_{\psi}(t_N) = \nabla_{\psi(t_N)} \mathfrak{L}(\{\ket{\psi(t_i)}\}) \,,
\end{align}
where the standard conversion factor $C^{\rm IS}_{\psi(t)}$
in Eq.~\eqref{Eq:adjoint_SDE} accounts for the required transformation from the It\^o to the Stratonovich sense and vice versa~\cite{kloeden2013numerical}. 
The gradients of the loss function $\vec{a}_{\vec{\theta}}(t_0) = \nabla_{\vec{\theta}}\mathfrak{L}$ with respect to $\vec{\theta}$ are then determined by the integration of
\begin{align}
     \der{\vec{a}_{\vec{\theta}}}(t)=& -\left(\vec{a}_{\psi}^\dagger(t) \cdot \nabla_{\vec{\theta}}\right) \left(K_{\psi(t)} - 2C^{\rm IS}_{\psi(t)}\right) \der{t}-\left(\vec{a}_{\psi}^\dagger(t) \cdot \nabla_{\vec{\theta}}\right) M_{\psi(t)}  \der{W(t)}\, ,
\label{Eq:adjoint_SDE2}
\end{align}
with the initial condition $\vec{a}_{\vec{\theta}}(t_N)= \vec{0}_{{\rm Dim}[\theta]}$.
This continuous adjoint sensitivity method  will be used in Section~\ref{Subsec:Continuous}.
Further details regarding the adjoint method and our Julia implementation~\cite{GSOC,bezanson2012julia} within the SciML ecosystem~\cite{rackauckas2020universal,rackauckas2019diffeqflux,rackauckas2017differentialequations} are discussed in \ref{App:Implementation}.

\section{SDE control based on full knowledge of the state}
\label{Sec:SDE_control_state}

We first investigate the scenario in which the controller maps the quantum state $\ket{\psi(t)}$ to a new control parameter, $\mathcal{C}: \ket{\psi(t)} \mapsto \Omega(t) \in [ -\Omega_{\rm max}, \Omega_{\rm max}]$.
From a learning perspective, this is a major simplification because the NN does not need to learn how to filter the  homodyne current $J(t)$ to determine the state $\ket{\psi(t)}$. 
From a practical perspective of controller design, this approach assumes that there is already a filter module, which allows one to predict the state of the system from the measurement record and past control actions. We discuss in Section~\ref{Subsec:Homodyne} and \ref{app: filter} the implementation of such a filter in our case of a qubit subjected to homodyne detection. Note that if the initial state of the qubit is unknown, only a mixed state $\rho_t$ can be obtained if the detection record is too short. 
Nevertheless, it is always possible to obtain a pure state estimate $\rho_t\mapsto \ket{\psi(t)}$ and recover our scenario, e.g.\ by projecting $\rho_t$ onto the surface of the Bloch sphere.
Alternatively, one can straightforwardly generalize our approach to the case of a stochastic quantum master equation describing the evolution of the system's density matrix in the presence of homodyne detection, cf.~\ref{App:HD:DerivationNormPreservingSDE}.

For all numerical experiments we fix the parameters of the physical model as $\Delta=20 \kappa$ and $\Omega_{\rm max}=10 \kappa$. 
We discuss the variation of these parameters and their impact on the reached fidelity in \ref{App:kappa}.

\subsection{Hand-crafted strategy}
\label{Subsec:Hand-crafted}

\begin{figure}[h!]
	\centering
	\includegraphics[width=1.0\linewidth,angle=0]{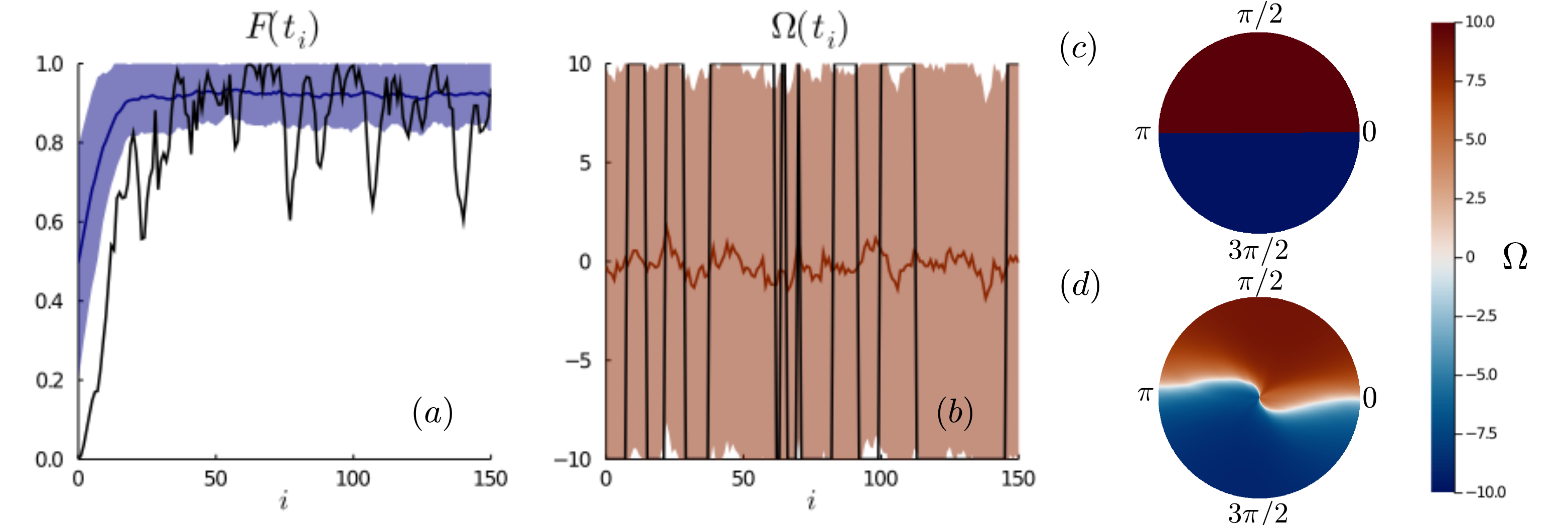}
	\caption{
	Preparation of the state $\ket{e}$ for the continuously monitored qubit described by the SDE~\eqref{Eq:SDE_homodyne}.
	(a) Mean fidelity (blue solid line) and corresponding standard deviation (shaded) for the hand-crafted strategy summarized in Algorithm~\ref{algo:handcrafted} as a function of the time steps $i$ at which checkpoints are stored.
	(b) Corresponding drive amplitudes.
    The black lines in (a,b) visualize the time evolution of the initial state $\ket{g}$.
    Stereographic projection, Eq.~\eqref{Eq:stereographic_projection}, of the applied drive $\Omega$ with respect to states $\ket{\psi}$ of the southern hemisphere of the Bloch sphere for (c) the hand-crafted strategy and (d) the optimized NN described in Section~\ref{Subsec:Continuous}.
    Note that panel (d) is the stereographic projection of the Bloch sphere of the inset of Fig.~\ref{Fig:4}(a).}
	\label{Fig:3}
\end{figure}

The control operator $\sigma_x$ in Eq.~\eqref{Eq:H} induces a rotation about the $x$-axis. 
Therefore, a simple but very intuitive strategy to move the state upwards in each time step is to compute the expectation value $\ave{\sigma_y}_{\psi(t)}$ and to choose the direction of rotation depending on its sign.
Specifically, if $\langle\sigma_y\rangle > 0$ (or $\langle\sigma_y\rangle<0$), the controller should rotate (counter-) clockwise about the $x$-axis, as summarized in Algorithm~\ref{algo:handcrafted}.

\begin{minipage}[t]{0.49\linewidth}
       \begin{algorithm}[H]
       \DontPrintSemicolon
       \SetAlgoLined
       \KwIn{$\psi(t)$}
       \KwResult{$\Omega(t)$}
       \eIf{$\langle \sigma_y \rangle_{\psi(t)} > 0$}{\vspace{1.0mm} $\Omega(t) \leftarrow \Omega_{\rm max}$\;}{$\Omega(t) \leftarrow -\Omega_{\rm max}$\;}
       \caption{Hand-crafted controller}
       \label{algo:handcrafted}
       \end{algorithm}
\end{minipage}\vspace{0.5cm} 

Figure \ref{Fig:3}(a,b) shows the fidelity and the associated drive $\Omega$ for this control scheme. 
Though conceptually straightforward, this hand-crafted control function is very efficient.
The mean fidelity over the whole control interval is $F_{\rm hc}=0.90\pm 0.13$.
We visualize the control strategy in Fig.~\ref{Fig:3}(c), where we show the applied drive $\Omega$ as a function of the state on the Bloch sphere. 
The Bloch sphere [with spherical coordinates ${\vartheta}, \phi$, see also Eq.~\eqref{Eq:initial_state}] is mapped onto the tangential plane at $z=0$, described by polar coordinates $(R,\Phi)$, using a  stereographic projection
\begin{align}
   (R, \Phi) &= (\cot \frac{{\vartheta}}{2}, \phi) \,.
    \label{Eq:stereographic_projection}
\end{align}
Apparently, the stereographic projection maps the south pole to $R=0$ and the north pole to $R=\infty$.
Throughout this paper we truncate the value of $R$ to an interval $[0,1]$, so we plot the applied drive values on the states of the southern hemisphere, cf. Fig.~\ref{Fig:3}(c,d).

\subsection{Continuously updated control drive}
\label{Subsec:Continuous}

\begin{figure}[h!]
	\centering
	\includegraphics[width=1\linewidth,angle=0]{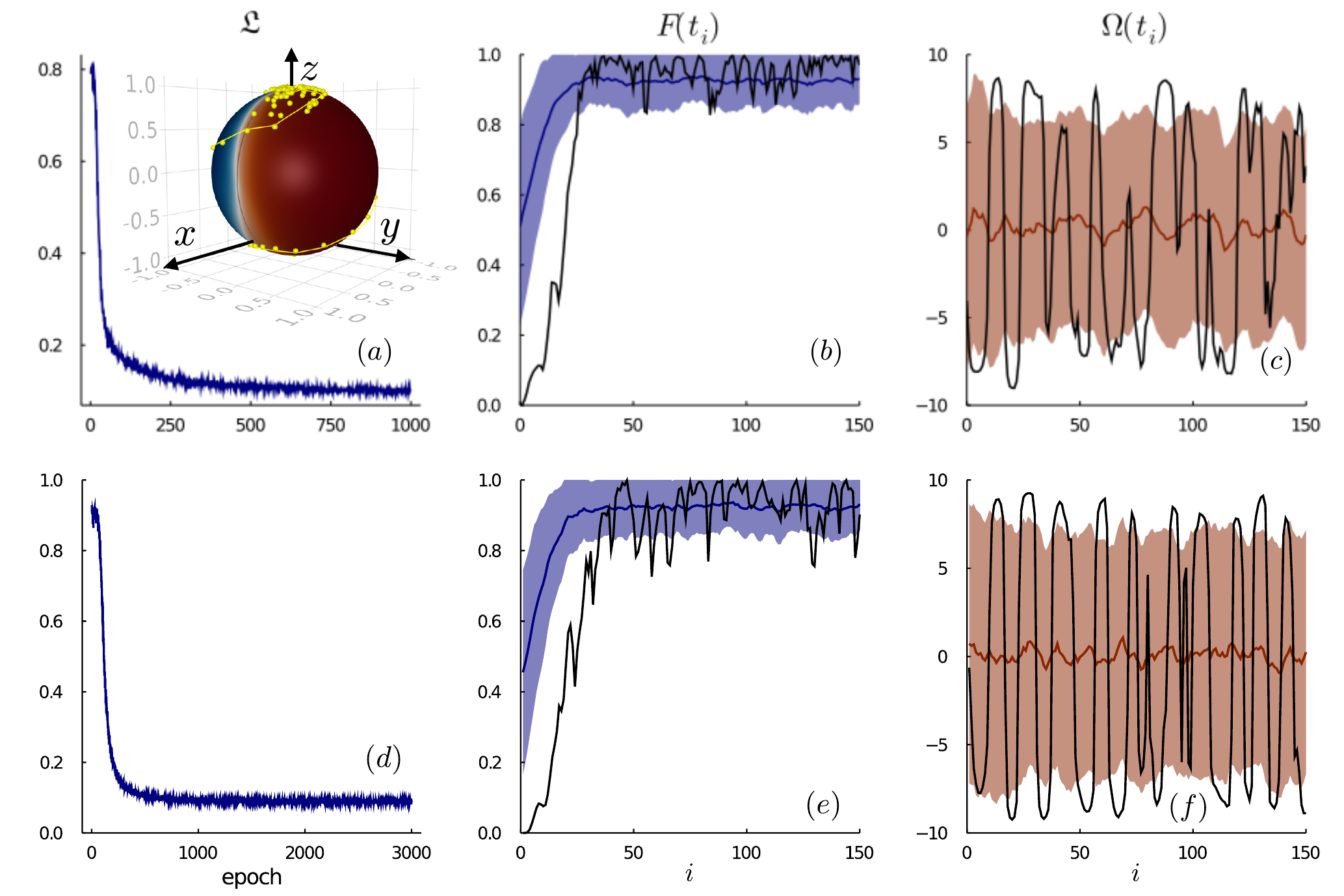}
	\caption{
	Preparation of the state $\ket{e}$ for the continuously monitored qubit described by the SDE~\eqref{Eq:SDE_homodyne}, based on (a-c) continuously updated control parameters and (d-f) piecewise-constant control parameters.	We compute the gradients of the loss function with respect to the parameters of the NN using the continuous adjoint sensitivity method in (a-c) and the nested discrete AD method in (d-f), see Section~\ref{Subsec:Adjoint}.
	(a,d) Smooth evolution of the loss as a function of training epochs.
	(b-f) Performance of the trained NN when it is applied to a set of 256 randomly sampled initial states on the Bloch sphere. 
	The solid blue [red] line in (b,e) [(c,f)] shows the mean fidelity $F(t_i)$ [control drive $\Omega(t_i)$] as a function of the time steps $t_i$ at which checkpoints are stored.
	The shaded regions represent the corresponding standard deviation which is  large in panels (c) and (f) because the controller will choose a different sequence of controls for each quantum state. 
    The black lines in (b-f) visualize the results for an initial state $\ket{g}$.
    Inset in Panel (a): Visualization of the control parameters $\Omega$ (in color code) as a function of the current state $\ket{\psi(t)}$ on the Bloch sphere. 
    The color palette indicates rotation clockwise (blue) or anti-clockwise (red) about the $x$-axis.
    The yellow points indicate the trajectory starting at the initial state $\ket{g}$ where the first 50 points are connected by a line.
    The hyperparameters are listed in Tab.~\ref{Tab:HP}.
    }
	\label{Fig:4}
\end{figure}

Now we apply a NN as the controller.
We use a controller which changes $\Omega(t)$ in every time step of the solver based on the current state $\ket{\psi(t)}$.
This implements a high-frequency feedback loop but renders discrete AD methods very inefficient since all parameters of the NN contribute to the feedback loop. 
Consequently, we use the continuous adjoint sensitivity methods described in Section~\ref{Subsec:Adjoint} to calculate gradients of the loss function.

Figure~\ref{Fig:4}(a) shows the smooth evolution of the loss function throughout the training of the (fully-connected) neural network, which converges to a configuration able to reach quickly fidelities with mean value of about $0.9$ with modest standard deviation, see Fig.~\ref{Fig:4}(b). 
The mean fidelity over the whole control interval is $F_{\rm c}= 0.90\pm 0.13$. 
Figure~\ref{Fig:4}(c) illustrates the applied drive $\Omega$ during this time evolution.
The inset of Fig.~\ref{Fig:4}(a) visualizes the control strategy on the Bloch sphere.
The same is depicted in Fig.~\ref{Fig:3}(d) using the stereographic projection, see Eq.~\eqref{Eq:stereographic_projection}.
The controlled evolution of the initial state $\ket{\psi(t_0)}=\ket{g}$, corresponding to the black lines in Fig.~\ref{Fig:4}(b,c), is marked by the yellow points.
These points show how the controller first transfers the state from the south pole to the north pole region and then stabilizes it in vicinity of the target state $\ket{e}$.

\subsection{Piecewise-constant control drive}
\label{Subsec:Piecewise}

Now we reduce the control frequency and assume that the controller changes the action $\Omega(t)$ only every $N_\mathrm{sub}$ time steps.
This case is crucial in many physical situations, where the control loop is not fast enough to follow high-frequency changes in the physical system.
In practice this means that we evolve the state for $N_{\rm sub}$ substeps between the checkpoints $\{\ket{\psi(t_i)},\Omega(t_i) \}$ with fixed value of $\Omega(t_i)$.
The resulting piecewise-constant control scheme allows us to use the discrete forward-mode adjoint sensitivity method through the SDE solver combined with an outer reverse-mode AD through the rest of the control loop, as described in Section~\ref{Subsec:Adjoint}.
Although we restricted the rate at which $\Omega(t)$ can change, we find that the NN converges to a similar learning behavior with similarly large fidelities $F(t)$ as in Section~\ref{Subsec:Continuous}, see Fig.~\ref{Fig:4}(e).
The mean fidelity over the whole control interval is $F_{\rm pw}=0.89\pm 0.10$. 

\subsection{Comparison of the control strategies}
\label{Subsec:Comp}
Based on the results from a set of 256 trajectories, all three 
control strategies perform nearly equally well in terms of their average fidelities F $\approx$ 0.9.
The piecewise-constant controller slightly outperforms the other two approaches by having the smallest relative dispersion of the mean fidelity $F_{\rm pw}$, which we attribute to the larger number of training epochs and the larger NN, see Table~\ref{Tab:HP}. 

The hand-crafted strategy achieves a large average fidelity, but generates sudden jumps in the drive $\Omega(t)$, as shown in Fig.~\ref{Fig:3}(b). Such a drive is experimentally hardly feasible.
We find that NNs with moderate depths provide a smooth mapping between the input states and the drive, see Figs.~\ref{Fig:3}(c) vs.~\ref{Fig:3}(d), while keeping high fidelity in the control interval.
The signals generated by these protocols are experimentally more accessible, see Figs.~\ref{Fig:4}(d) and~(f).
When necessary, specific terms can be added to $\mathfrak{L}$ in Eq.~\eqref{Eq:Loss} to strengthen various requirements on the controller's performance (e.g.\ the smoothness of the drive or bounds on the power input) as discussed in Section~\ref{Subsec: control}. 
This is not the case for the hand-crafted strategies where an efficient implementation of these constraints might be impossible. 
Furthermore, in some cases, e.g. the mountain car problem, it is not straightforward to come up with any hand-crafted strategy to start with. 
In contrast, the RL and \DP{} approach is easily adjustable to different physical systems.

There are two principal reasons why the fidelity $F$ only reaches 0.9 in our setup. 
First, $F$ is a time-average and the controller needs some time to align the qubit to a target state (starting from an arbitrary state).
Second, the controlled qubit rotates about an axis in the $x$-$z$-plane whose direction depends on the ratio $\Delta/\Omega$. 
Hence, even in the case $\Omega \gg \Delta$ when the qubit rotates solely about the $x$-axis, the drive can only bring the qubit all the way up to the north pole of the Bloch sphere if the current state lies on a great circle perpendicular to this axis.
Therefore, the ratio $\Delta/\kappa$ and the maximum value of $\Omega$ set a limit on how close the qubit can come to the target state on average.
In \ref{App:kappa}, we study scenarios with lower $\kappa$ and observe that the average fidelity can increase and approach unity. We can thus conclude that $F\approx 0.9$ is not a limit of our design but rather originates from the restricted capability of control operations considered here.

\section{SDE control based on homodyne current}
\label{Sec:SDE_control_Jhom}

\begin{figure}[h!]
	\centering
	\includegraphics[width=1\linewidth,angle=0]{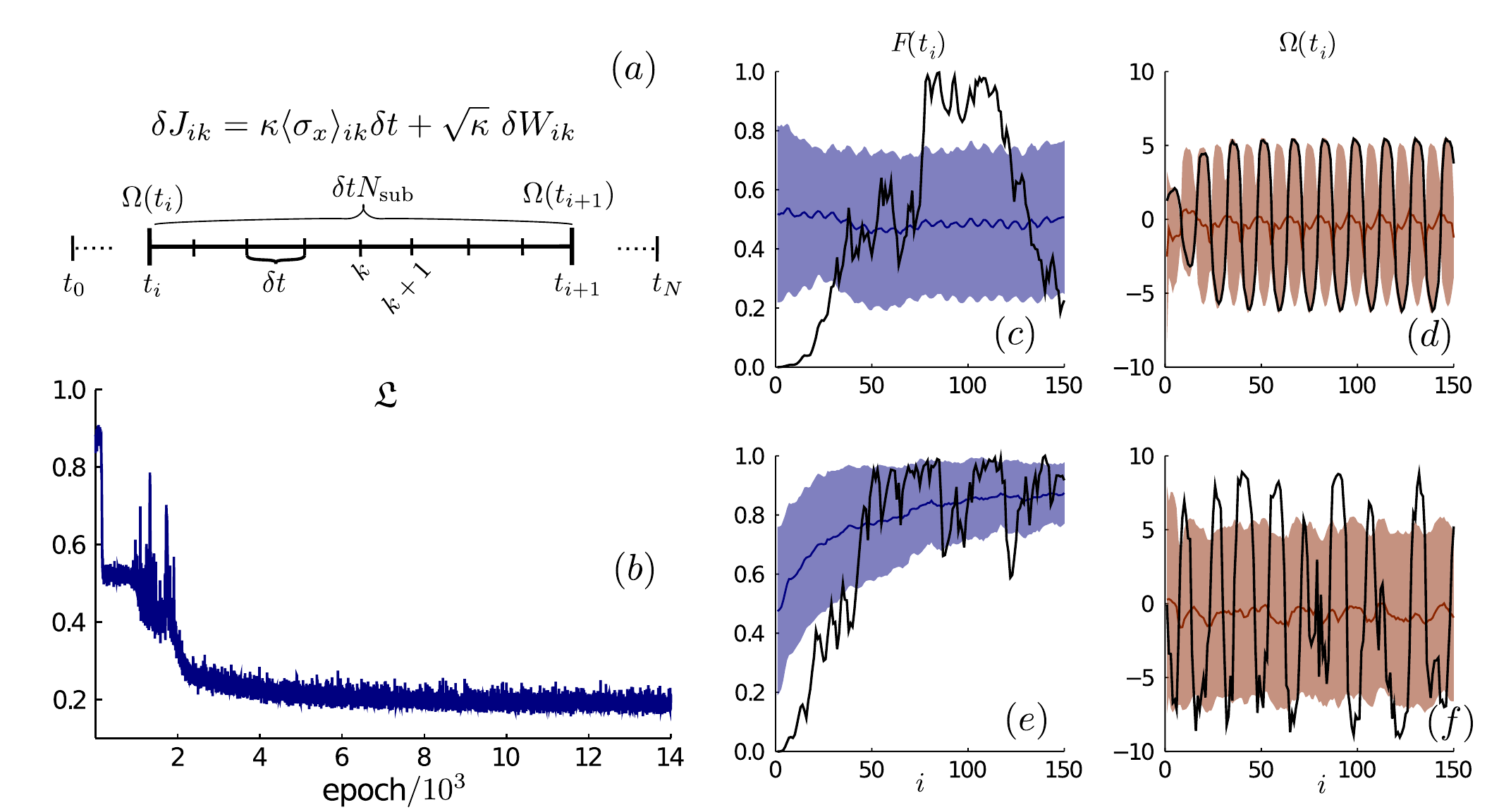}
	\caption{Preparation of the state $\ket{e}$ using the piecewise-constant control scheme with the discrete AD method to compute the gradients $\nabla_{\vec{\theta}}\mathfrak{L}$, see Section~\ref{Subsec:Adjoint}, where only the homodyne current $J(t)$ is used as an input to the controller.
	(a) Sketch of the acquisition process of the input data for the NN controller, see Section~\ref{Sec:SDE_control_Jhom}.
	(b) Evolution of the loss function $\mathfrak{L}$ as a function of the training epochs.
	(c) and~(d) Performance of the NN after 400 training epochs [in the first plateau of $\mathfrak{L}$ in panel (b)] for a set 256 trajectories: (c) Mean fidelity  (blue solid lines) and corresponding standard deviation (shaded area), (d) Corresponding drive amplitudes.
	(e) and (f) Same quantities as in panels (c) and (d), but for the fully trained NN, i.e.\ after 14000 training epochs.
	Black lines in panels (c-f) show the results for an initial state $\ket{g}$.
    The hyperparameters are listed in Tab.~\ref{Tab:HP} in \ref{App:HP}.}
	\label{Fig:5}
\end{figure}

\begin{minipage}[t]{0.49\linewidth}
       \begin{algorithm}[H]
       \DontPrintSemicolon
       \SetAlgoLined
       \KwIn{$\psi(t_0)$, $t_0$, $\Omega(t_0)=0$}
       \KwResult{$\mathfrak{L}$}
        \For{$i=0:N-1$}{
           \emph{compute and store checkpoints}\\
            $\Omega(t_{i+1}) \leftarrow \text{NN}_{\vec{\theta}}({ \{\bm{J}_\tau(t),\bm{\Omega}_m(t)\}})$\\
           $\psi(t_{i+1}) \leftarrow $ solve($\psi(t_i)$,$\Omega(t_{i+1})$) for SDE \eqref{Eq:SDE_homodyne} in $[t_i, t_{i+1}]$
        }
        $\mathfrak{L}  \leftarrow$ loss($\{\psi(t_i), \Omega(t_i)\}$)
       \caption{Homodyne current}
       \label{algo:homodyne}
       \end{algorithm}
\end{minipage}\vspace{0.5cm} 

In this section, we will construct a controller which directly obtains the (noisy) measurement record of the homodyne current and determines the optimal control field $\Omega(t_i)$ in each time interval $[t_i,t_{i+1}]$.
We consider a controller formed by a slightly augmented NN architecture with fully connected layers (see~Fig.~\ref{Fig:Jhom_NN}).
The acquisition of input data for the NN consists of the following steps:
\begin{enumerate}
    \item  The controller generates a piecewise-constant drive $\Omega(t_i)$ between two checkpoints at times $t_i$ and $t_{i+1}$, as described in Section~\ref{Subsec:Piecewise}.
    In the time window $[t_{i},t_{i+1}]$, we integrate the SDE~\eqref{Eq:SDE_homodyne} using $N_{\rm sub}$ substeps of length $\delta t$ as sketched in Fig.~\ref{Fig:5}(a).
    We label these substeps with $k$.
    The input for the NN to predict the next $\Omega(t_{i+1})$ is a vector $\bm{J}_\tau(t_i)=[\delta J_{i1},\ldots,\delta J_{iN_{\rm sub}}]^T$ where $\tau=N_{\rm sub} \delta t$ is the length of the time interval over which we gather the data.
    Experimentally, $\delta t$ can be interpreted as the detection time window of the photodetectors. 
    According to Eqs.~\eqref{Eq:Jhom} and~\eqref{Eq:Jhom2}, the homodyne measurement in the $k$th substep is
    \begin{equation}
        \delta J_{ik}=\kappa \ave{\sigma_x}_{ik} \delta t + \sqrt{\kappa} \ \delta W_{ik}\,.
    \end{equation}
    The first term corresponds to the quadrature signal $\ave{\sigma_x}_{ik}$ measured over the detection window $\delta t$, which we assume to be approximately constant on  time scales of the order of $\delta t$.
    The second term $\delta W_{ik} \equiv  W_{i(k+1)}-W_{ik}$ is the Wiener increment during the $k$th substep.
     Note that the quantity $\delta J_{ik}$ is dimensionless and solely represents the number of measured detected photons, as discussed in~\ref{App:Homodyne}.
   
    \item In addition, we provide the NN  with the information on the $m$ last control parameters $\bm{\Omega}_{m}(t_i)= [\Omega(t_{i-1}), \ldots, \Omega(t_{i-m})]^T$.
    This equips the NN with a ``memory'' of its own actions, such that it can take into account how a sequence of the last control drive amplitudes affected the performance.
    We empirically choose $m$ such that the length of the input vector $\bm{\Omega}_m(t_i)$ corresponds to $1/10$ of the length of $\bm{J}_\tau(t_i)$.
\end{enumerate}

Given these inputs, the task for the controller is to provide an optimal mapping $\mathcal{C}: \{\bm{J}_\tau(t),\bm{\Omega}_m(t)\} \mapsto \Omega(t) \in [ -\Omega_{\rm max}, \Omega_{\rm max}]$.
Figure~\ref{Fig:5}(b) shows the evolution of the loss function $\mathfrak{L}$ during the learning phase as a function of the training epochs. 
Note that there are two distinct plateaus.
First, after a couple of hundred epochs, the NN develops a general strategy consisting of the application of a periodic drive to prevent the qubit from decaying to the ground state.
Figures~\ref{Fig:5}(c,d) show examples of the performance of the NN at epoch $400$ to illustrate this phase of the training.
This strategy is state-independent, yet it manages to keep the mean fidelity of the simulated trajectories around $0.5$ over the whole control interval.
Around epoch $2500$, after some transition phase where $\mathfrak{L}$ oscillates substantially, the NN starts to provide state-sensitive control fields.
The final performance of the controller in Fig.~\ref{Fig:5}(e,f) reaches the mean fidelity  $F_J=0.79 \pm 0.17$  over the whole control interval. 
During the last 50 time steps in the control interval, which we specifically focus on during the training using the adjusted loss function term of the type~\eqref{Eq:Infidelity} for these steps, the average fidelity is $F_J^{50}=0.86 \pm 0.12$.

At this stage, we infer that the NN has learnt how to extract the signal from the noisy data before it attains a similar learning strategy as seen in Section~\ref{Sec:SDE_control_state}.
In contrast to the filter mentioned in Section~\ref{Subsec:Homodyne} and \ref{app: filter}, this universal approach based on \DP{} will also work if some parameters of the model were a priori unknown. 
For example, if the detuning $\Delta$ was unknown, one could train the controller on an ensemble of $M$ randomly chosen parameters $\{\Delta_k\}_{k=1}^M$, such that it learns how to deal with the general situation of arbitrary detuning. 
In this case a straightforward filtering of the signals to obtain the state is unfeasible, as this would require to solve the filter for all possible values of model parameters, cf.\ Eq.~\eqref{eq: filter app}.
Other options for signal filtering would be (recursive) backward filtering methods~\cite{schauer2017, schauer2020automatic} or recurrent neural networks because their structure allows them to capture temporal correlations in the data~\cite{flurin2020}.
Note that such filter methods are compatible with the two-step control approach described in Section~\ref{Sec:SDE_control_state}.

\section{Discussion}
\label{Sec:Conclusion}

In this work we proposed a framework based on differentiable programming (\DP{}) to automatically design feedback control protocols for stochastic quantum dynamics.
As a test bed, we used a qubit subjected to homodyne detection, whose dynamics is given by a stochastic Schr\"odinger equation.
Note, however, that our method can straightforwardly be applied to different physical systems and that it can be generalized to the case of stochastic quantum master equations.

In Section~\ref{Sec:SDE_control_state}, we demonstrated that a controller, formed by a NN, can be trained to prepare and stabilize a target state starting from an arbitrary initial state of the system when the NN obtains the full knowledge of the instantaneous state at any time.
The method generates a smooth drive $\Omega(t)$ while maintaining a high fidelity with the target state over the whole control interval.
Additional constraints on the performance of the controller can be implemented by adding further terms to the loss function.
This makes the \DP{} approach more versatile than tailoring control functions manually, which requires a unique approach for every new system and can even be infeasible for large quantum systems.

The key feature of our \DP{} framework is to include an explicit model of the system's dynamics into the computation of the gradients of the loss function with respect to the parameters of the NN, i.e.\ the controller.
Specifically, in Section~\ref{Subsec:Continuous}, we employed the recently developed continuous adjoint sensitivity method for the gradient estimation through an SDE solver, which is memory efficient and, thus, allows us to study a high-frequency controller.
F.S. implemented these new continuous adjoint sensitivity methods in the DiffEqSensitivity.jl package within the open-source SciML ecosystem~\cite{GSOC}.

In Section~\ref{Sec:SDE_control_Jhom}, we showed that the feedback control scheme can be based directly on providing the NN with a record of homodyne current measurements without the need to filter the information on the actual state beforehand.
Therefore, the NN must first learn how to filter the input data (with poor signal-to-noise ratio) before it can predict optimal state-dependent values of the control drive.
Ultimately, the trained NN was able to reach fidelities above $85\%$ in a target time interval for random initial states.

In future studies, the optimization of the loss function based on stochastic trajectories using adjoint sensitivity methods could be compared to alternative approaches. 
First, the solution to the stochastic optimal control problem in the specific case of Markovian feedback (as in Section~\ref{Sec:SDE_control_state}) is a Hamilton-Jacobi-Bellman equation~\cite{rackauckas2020universal,kloeden2013numerical}. The solution of this partial differential equation, with same dimension as the original SDE, may directly give the optimal drive~\cite{Gough2005}.
However, solving this partial differential equation with a mesh-based technique is computationally demanding and mesh-free methods, e.g. based on NNs also require a (potentially costly) training procedure~\cite{sirignano2018dgm}.
Second, the expected values of the loss functions could be optimized by leveraging the Koopman expectation for direct computation of expected values from stochastic and uncertain models~\cite{gerlach2020koopman}.
Additionally, one could approach this control problem by using an SDE moment expansion to generate ordinary differential equations for the moments and apply a closure relationship \cite{lamperski2018analysis}. 
Additional research is required to ascertain the efficiency of these approaches in comparison to our method.

The results reported in this paper imply that \DP{} is a powerful tool for the automated design of quantum control protocols.
Further experimental needs, e.g.\ finite time lag between the measurement and the applied drive, finite-temperature effects, or imperfect homodyne detection, can be incorporated straightforwardly into this method.
Thus, our work introduces a new perspective on how prior physical knowledge can be encoded into machine learning tools to construct a universal control framework.
Besides the control application demonstrated here, the \DP{} paradigm can be also adopted to solve other inverse problems such as estimating model parameters from experimental data~\cite{greplova2017quantum,valenti2019hamiltonian,krastanov2019}. 
An interesting perspective for future work is to extend our framework to control-assisted quantum sensing and metrology.

\section*{Acknowledgment}
We would like to thank Niels L\"orch, Eliska Greplova, Moritz Schauer, and Chris Rackauckas for helpful discussions. 
We acknowledge financial support from the Swiss National Science Foundation (SNSF) and the NCCR Quantum Science and Technology. 
Parts of the computations were performed at sciCORE (scicore.unibas.ch) scientific computing core facility at University of Basel.

\section*{Data availability statement}
The codes that support the findings of this study are openly available~\cite{CODE}.

\appendix
\section{Continuous homodyne detection}
\label{App:Homodyne}

\subsection{Quantum trajectories: monitoring the spontaneous emission of a qubit}

Consider a two-level atom  interacting with a free photon.
In the case of discrete photon modes, the interaction is given by the usual Hamiltonian $\bar{H}_{\rm int}= \ii g (\hat{\sigma}_+ \hat{a} - \hat{\sigma}_- \hat{a}^\dag)$ with $\hat{a}^\dagger$ and $\hat{a}$ being the bosonic creation and annihilation operators, respectively, fulfilling the commutation relation $[\hat{a},\hat{a}^\dag]=1$, and $g$ being a coupling constant with dimension of energy.
In contrast to the notation used in the main text, in this appendix, we will mark operators by a hat to distinguish them from scalars.
This Hamiltonian results from the dipole interaction written in the form $\bar{H}_{\rm int}\propto(\hat{a} + \hat{a}^\dag) \hat{\sigma}_y$ and application of the rotating wave approximation.
However, when addressing a decay to the continuum, the photons are represented by quantum field operators with commutation relation $[\hat{a}_t^{\phantom{\dag}},\hat{a}_s^\dag]=\delta(t-s)$, thus having units of square root of energy (or, equivalently, inverse square root of time).
The interaction Hamiltonian takes the form
\begin{equation}
 \hat{H}_{\rm int} = \ii \sqrt{\kappa}(\hat{\sigma}_+ \hat{a}_t - \hat{\sigma}_- \hat{a}_t^\dag)\,,
 \label{EqAp:Hint}
\end{equation}
where $\kappa$ is the decay rate, again, with dimension of energy like $g$.

The field propagator in the interaction picture can be formally written as $\hat{U}_\tau =\mathcal{T}\left[\mathrm{e}^{-\ii \int_t^{t+\tau} \hat{H}_\t{int}(s)\der{s}}\right]$, where $\mathcal{T}$ is the time-ordering operator.
Expanding $\hat{U}_{\delta t}$ for short times $\delta t$, we find
\begin{equation}
\begin{split}
    \hat{U}_{\delta t} = &1 + \sqrt{\kappa}\int_t^{t+\delta t} \!\! (\hat{\sigma}_+ \hat{a}_s - \hat{\sigma}_- \hat{a}_s^\dag) \der{s}\\
    &-\frac{\kappa}{2}\int_t^{t+\delta t} \!\!( \hat{\sigma}_- \hat{\sigma}_+  \hat{a}_s^\dag \hat{a}_{s'}^{\phantom{\dag}} + \hat{\sigma}_+\hat{\sigma}_-  \hat{a}_s^{\phantom{\dag}} \hat{a}_{s'}^\dag ) \der{s} \der{s}'.
\end{split}
\label{EqAp:ShortTimeProp}
\end{equation}
Here, the second-order terms give an important $O(\delta t)$ contribution. Higher-order terms only contribute as $O(\delta t^{3/2})$ and can be safely neglected, and we also assume that $\delta t$ is short on the timescale of internal qubit dynamics.  
Assuming the free field is originally in the vacuum state $\ket{0}$ we can write
\begin{equation}
    \hat{U}_{\delta t} \ket{0} = \Big(1 - \frac{\kappa \delta t}{2} \prjct{e}\Big)\ket{ 0} + \sqrt{\kappa \, \delta t} \, \hat{\sigma}_- \ket{ 1}_t, 
    \textnormal{where} \ \ket{ 1}_t =  \frac{1}{\sqrt{\delta t}} \int_{t}^{t+\delta t} \!\! \hat{a}_s^\dag \der{s} \ket{ 0}\,
\end{equation}
is a properly normalized state of the field, and $\ket{e}$ denotes the excited qubit state.

The simplest way to monitor the qubit state $\ket{\psi}$ is then to continuously measure the intensity of the outgoing field.
This defines a Poissonian process, as the probability to detect a photon in a time interval $\delta t$ reads 
\begin{align}
    p_1 
    &= \tr_\text{qubit}\left\lbrace \bra{1} \hat{U}_{\delta t} \prjct{0}\otimes \prjct{\psi} \,  \hat{U}_{\delta t}^\dag \ket{ 1} \right\rbrace \nonumber \\
    &=  \kappa \delta t \, \tr \left\lbrace \hat{\sigma}_- \prjct{\psi} \hat{\sigma}_+ \right\rbrace 
    = \kappa \delta t \abs{\scal{e}{\psi}}^2,
\end{align}
and the probability of no detection is $p_0=1-p_1$ respectively.
Thus, we obtain the standard unraveling for the spontaneous decay process
\begin{equation}
\begin{split}
    \text{one photon emitted}&:\qquad \sqrt{\kappa \delta t} \hat{\sigma}_- \ket{\psi}, \\
    \text{no photons emitted}& :\qquad \Big(1 - \frac{\kappa \delta t}{2} \prjct{e}\Big)\ket{\psi} .
\end{split}
\end{equation}

\subsection{Weak homodyning}

An alternative to photon-counting detection is to mix the outgoing light with coherent light from a local-oscillator laser on a beamsplitter, and to measure the intensities at both output ports of the beamsplitter.
As a warm-up we first consider the situation where the intensity of the local-oscillator field is comparable to the emitted light.
It is given by the expansion of a coherent state using only vacuum and single-photon states
\begin{equation}
\ket{\sqrt{\delta t} \beta} = \left(1- \frac{|\beta|^2}{2} \delta t\right)\ket{0}  +\beta \sqrt{\delta t} \ket{1},
\end{equation}
where the single-photon state of the $\hat{b}$-mode is defined identically to the mode $\hat{a}$. 
Mixing the two states on a 50:50 beamsplitter gives
\begin{equation}
\label{eq: weak homo}
\begin{split}
    \hat{U}_{BS} \hat{U}_{\delta t} \ket{0} \ket{\sqrt{\delta t} \beta } 
    = &\left(1-\frac{\kappa \delta t} {2}\prjct{e} - \frac{|\beta|^2 \delta t}{2}\right)\ket{ 0,0}\\
    & + \sqrt{\frac{\delta t}{2}} \big(\beta +\sqrt{\kappa} \hat{\sigma}_-\big) \ket{ 1,0}\\
    & + \sqrt{\frac{\delta t}{2}} \big(\beta -\sqrt{\kappa} \hat{\sigma}_-\big) \ket{ 0,1}.
\end{split}
\end{equation}
Introducing the photocurrent variable $q=n-m$, which measures the intensity difference of the two outputs, we see that there are three possibilities, $q=-1,0,1$. The respective probabilities of the three outcomes are given by the norms of the three different terms in Eq.~\eqref{eq: weak homo},
\begin{equation}
\begin{split}
    p_{q=\pm 1} &= \frac{\delta t}{2} \ave{ \beta^2 \pm \beta \sqrt{\kappa} \hat{\sigma}_x + \kappa \prjct{e}}_{\psi}\,,\\
     p_{q=0} &= 1- \delta t \beta^2 - \delta t  \kappa \ave{\prjct{e}}_{\psi}\,.
\end{split}
\end{equation}
 The three possible states of the system after the measurement are proportional to
\begin{equation}
\begin{split}
q=\pm 1 &:\quad \big(\beta \pm \sqrt{\kappa} \hat{\sigma}_-\big) \ket{\psi},\\
q=0 &: \quad \left(1-\frac{\kappa \delta t} {2}\prjct{e} - \frac{|\beta|^2 \delta t}{2}\right)\ket{\psi}.
\end{split}
\end{equation}
By setting $\beta=0$ (i.e. no mixing with a local-oscillator field), we recover the previous case where the qubit simply has a chance to decay
\begin{equation}
q=\pm 1 : \qquad \sqrt{\kappa}\hat{\sigma}_- \ket{\psi},
\end{equation}
and $p_{q=1}+p_{q=-1}=p_1$. The only difference is that the emitted photon is randomly split between the two detectors, with no information about the state of the qubit contained in the sign of $q$. 

\subsection{Strong homodyning}

Finally, we  consider the situation treated in the main text. 
The standard homodyne measurement requires mixing the signal with a strong coherent local-oscillator field $\abs{\beta} \gg 1$ 
\begin{equation}
\ket{\beta}=  \sum_{n=0}^\infty c_n(\beta) \ket{n}, \ 
c_n(\beta)  = \mathrm{e}^{-\beta^2/2} \frac{\beta^{n}}{\sqrt{n!}}.
\end{equation}
Recall that here the Fock states $\ket{n}= \frac{\hat{b}^{\dag n}}{\sqrt{n!}}\ket{ 0}$ are defined with the creation operator $\hat{b}^\dag=\frac{1}{\sqrt{\delta t}}\int_t^{t+\delta t} \hat{b}_s^\dag\, \der{s}$, describing the field impinging on the lower port of the beamsplitter during the interval $\delta t$, see Fig.~\ref{Fig:1}(a). 
For the modes $\hat{a}$ and $\hat{b}$ to match, the local-oscillator laser $\beta$ has to be in resonance with the drive laser $\Omega$ because, in the lab frame, the field $a_t$ picks up the phase of $\hat{\sigma}_-$ rotating at the frequency of the drive laser. 

When the modes match, the 50:50 beamsplitter transformation takes the form
\begin{equation}
\hat{U}_{BS}\binom{\hat{a}^\dag}{\hat{b}^\dag} = \binom{\frac{\hat{a}^\dag-\hat{b}^\dag}{\sqrt{2}}}{\frac{\hat{a}^\dag+\hat{b}^\dag}{\sqrt{2}}}.
\end{equation}
In particular, this implies $\hat{U}_{BS}\ket{0,\beta} = \ket{\frac{\beta}{\sqrt{2}},\frac{\beta}{\sqrt 2}}$, i.e.  a coherent state $\ket{\beta}$ of the mode $b$ is split into two coherent states of half the intensity when mixed with the vacuum state $\ket{0}$ of the mode $a$ on a 50:50 beamsplitter.
Using the last two expressions, we can write down the overall state of the qubit, the spontaneously emitted radiation and the homodyne laser field after the beamsplitter. To do so, let us compute
\begin{equation}\label{eq: U tilde 1}\begin{split}
    \tilde{\mathds{U}}_{\delta t} &= (\hat{\id}_\t{qubit}\otimes \hat{U}_{BS}) (\hat{U}_{\delta t}\otimes \hat{\id}_\t{laser})\ket{0}\ket{\beta}\\  
    &= (\hat{U}_{BS}\otimes \hat{\id}_\t{qubit}) \left( 1-\frac{\kappa \delta t}{2}\prjct{e}+ \sqrt{\kappa \delta t} \hat{a}^\dag \hat{\sigma}_- \right)\ket{0,\beta} \\
    &=\left( 1-\frac{\kappa \delta t} {2}\prjct{e} + \sqrt{\frac{ \kappa \delta t}{2}} (\hat{a}^\dag - \hat{b}^\dag) \hat{\sigma}_-\right) \ket{\frac{\beta}{\sqrt{2}},\frac{\beta}{\sqrt{2}}},
\end{split}
\end{equation}
which is the operator mapping the state of the qubit at time $t$ to the state of the qubit and the two detected modes at time $t+\delta t$. We can further simplify this expression by noting that 
\begin{equation}\begin{split}
    \hat{a}^\dag \ket{\beta} 
    &= \mathrm{e}^{-\beta^2/2}\sum_{n=0} \frac{\beta^n}{\sqrt{n!}} \hat{a}^\dag \ket{n} 
    = \mathrm{e}^{-\beta^2/2}\sum_{n=0} \frac{\beta^n}{\sqrt{n!}} \sqrt{n+1} \ket{n+1} \\
    &= \mathrm{e}^{-\beta^2/2}\sum_{n=0} \frac{\beta^{n+1}}{\sqrt{(n+1)!}}  \frac{n+1}{\beta} 
    = \mathrm{e}^{-\beta^2/2}\sum_{n=0} \frac{\beta^{n}}{\sqrt{n!}} \frac{n}{\beta} \ket{n}\\
    &= \frac{\hat{n}}{\beta} \ket{\beta},
\end{split}
\end{equation}
where $\hat{n} = \hat{a}^\dag \hat{a}$ is the photon number operator. Labeling the photon number operator for the $\hat{b}$ mode as $\hat{m} = \hat{b}^\dag \hat{b}$, we can then rewrite Eq.~\eqref{eq: U tilde 1}  in a very intuitive form
\begin{equation}\label{eq: final state homo}
    \tilde{\mathds{U}}_{\delta t} = \left( 1-\frac{\kappa \delta t} {2}\prjct{e} + \sqrt{\kappa \delta t}\,  \frac{\hat n -\hat m}{\beta} \hat{\sigma}_-\right)  \ket{\frac{\beta}{\sqrt{2}},\frac{\beta}{\sqrt{2}}}.
\end{equation}
Again, $ \tilde{\mathds{U}}_{\delta t} \ket{\psi_t}$ directly gives us the state of the qubit and the detected modes, while 
\begin{equation}
    \bra{n,m} \tilde{\mathds{U}}_{\delta t} \ket{\psi_t} = \sqrt{p_{n,m}} \ket{\psi_{t+\delta t}|_{n,m}}
\end{equation}
gives the conditional state of the qubit together with the probability to detect $n$ and $m$ photons respectively.
In the measurement process, superpositions of states with different photon numbers collapse such that the state after the post-measurement is a classical statistical mixture of the possible outcomes,
\begin{equation}
    \sum_{n,m=0}^{\infty} p_{n,m} \prjct{\psi_{t+\delta t}|_{n,m}}\otimes \prjct{n,m}.
\end{equation}

One easily sees from Eq.~\eqref{eq: final state homo} that $ \ket{\psi_{t+\delta t}|_{n,m}}= \ket{\psi_{t+\delta t}|_{n-m}}$, i.e.  only the difference $(n-m)$ of the photon counts reveals information on the state of the qubit, while the sum $(n+m)$ only describes the shot noise of the local oscillator. 
It is therefore sufficient to keep the difference $q=n-m$ and discard the sum, which defines the state
\begin{equation}
    \sum_{q=-\infty}^\infty p_{q} \prjct{\psi_{t+\delta t}|_q}\otimes \prjct{q}, \qquad \text{where} \qquad p_q=\sum_{n= \max(0,-q)}^\infty p_{n,n+q}.
\end{equation}
In a slight abuse of notation we can formally introduce the joint quantum state of the qubit and the count difference $q$ as $\hat{\mathds{U}}_{\delta t}\ket{\psi_t}$ with
\begin{equation}\label{eq: propag j}
    \hat{\mathds{U}}_{\delta t} 
    = \left( 1-\frac{\kappa \delta t} {2}\prjct{e} + \sqrt{\kappa \delta t}\,  \frac{\hat q}{\beta} \hat{\sigma}_-\right) \ket{ \tilde \Phi_\beta} 
    \quad \t{where} \qquad 
    \ket{\tilde \Phi_\beta} =\sum_{q=-\infty}^{\infty} \sqrt{\mu_j(\beta)} \ket{q},
\end{equation}
$\hat q\ket{q} =q\ket{q}$, which gives rise to the same post-measurement state. Here,
\begin{equation}\label{eq: j state discrete}
    \mu_q(\beta) = \sum_{n= \max(0,-q)}^\infty c_n^2(\beta/\sqrt{2}) c_{n+q}^2(\beta/\sqrt{2}) = \mathrm{e}^{-\beta^2} I_{q}(\beta^2),
\end{equation}
where $I_n(z)$ is the modified Bessel function of the first kind.

Next, we use that the distribution of $q$ on the right-hand side of Eq.~\eqref{eq: j state discrete} is well approximated by the normal distribution $\mathcal{N}(0,\beta^2)$ in the limit $\beta\gg 1$. Therefore, we can replace the state $\ket{ \tilde \Phi_\beta}$ of an integer-valued $q$ in Eq.~\eqref{eq: propag j} with
\begin{equation}\label{eq: Phi beta}
\ket{\Phi_\beta} = \int_{-\infty}^\infty \left[ \frac{1}{\sqrt{2  \pi} \beta} \exp(-\frac{q^2}{2 \beta^2})\right]^{1/2} \ket{q} \der{q}
\end{equation}
of a continuously valued and normally distributed $q$, and we also introduce a continuously valued operator $\hat q$. In the regime of interest $\beta\gg 1$ the actual value of $\beta$ does not play a role. To get rid of it, recall that $\beta^2 = I \delta t$ is the intensity of the local-oscillator laser in the time window $\delta t$, so it is more convenient to work with the laser power $I$, which is independent of the choice of the time window $\delta t$. Then, we can rescale the photon count difference to 
\begin{equation}
    j = \sqrt{\frac{\kappa}{I}} q
\end{equation}
to get rid of the laser intensity. 
Equation~\eqref{eq: propag j} now reads
\begin{equation}\label{eq: Udt j}
    \mathds{U}_{\delta t} =  \left( 1-\frac{\kappa \delta t} {2}\prjct{e} + \hat {j}\,  \hat{\sigma}_-\right) \ket{ \Phi},
\end{equation}
where the initial state $\ket{\Phi}$ of the rescaled $j$ can be obtained form Eq.~\eqref{eq: Phi beta} and satisfies
\begin{equation}
    P_0(j) =|\scal{j}{\Phi}|^2 = \frac{1}{\sqrt{2\pi \kappa \delta t}} \exp\left(-\frac{j^2}{2 \kappa \delta t}\right).
\end{equation}
This also allows us to define the homodyne current of the main text  $J = j/ \delta t $, as the photon count difference per time. At this point note that Eq.~\eqref{eq: Udt j} already implies  Eq.~\eqref{Eq:Homodyne_NoNorm}. Let us now  show how the measured value of $j$ is distributed. The marginal distribution of $j$ after the measurement reads 
\begin{equation}
\begin{split}
    P(j) 
    &= \norm{\bra{j} \hat{\mathds{U}}_{\delta t}\ket{\psi_t}}^2 
    = \norm{\bra{j} \left( 1-\frac{\kappa \delta t} {2}\prjct{e} + \hat {j}\,  \hat{\sigma}_-\right)\ket{\Phi} \ket{\psi_t}}^2 \\
    &= P_0(j) \norm{\left( 1-\frac{\kappa \delta t} {2}\prjct{e} + j\,  \hat{\sigma}_-\right)\ket{\psi_t}}^2\\
    &= P_0(j)\bra{\psi_t}\left( 1-\frac{\kappa \delta t} {2}\prjct{e} + j\,  \hat{\sigma}_+\right)\left( 1-\frac{\kappa \delta t} {2}\prjct{e} + j \,  \hat{\sigma}_-\right)\ket{\psi_t}\\
    & = P_0(j)\left(1 +(j^2 -\kappa \delta t)\ave{\hat{\sigma}_+\hat{\sigma}_-}_\psi  + j\ave{\hat{\sigma}_x}_\psi \right).
\end{split}
\end{equation}
From $P(j)$ one easily deduces the expected value and the variance of $j$ 
\begin{equation}
\mathds{E}(j) = \kappa \delta t \ave{\hat{\sigma}_x}_\psi\qquad \t{Var}(j)= \kappa \delta t .
\end{equation}
Clearly, it has the form of a Wiener process with drift
\begin{equation}
    j = \kappa  \ave{\hat{\sigma}_x}_\psi \delta t + \sqrt{\kappa}\, \delta W,
\end{equation}
where $\delta W$ is the increment of a Wiener process for an interval $\delta t$ (a normally distributed random variable with zero mean and variance $\delta t$). The last step is to include the Hamiltonian of the qubit $\hat{H}= \frac{1}{2}\left(\Delta \hat{\sigma}_z +\Omega(t) \hat{\sigma}_x \right)$ and set $\der{t} =\delta t$ such that $I^{-1}\ll \der{t} \ll \kappa^{-1}, \Delta^{-1}, \Omega^{-1}$ in order to  get
\begin{equation}\begin{split}
    \hat{\mathds{U}}_{\der{t}} &= \left(1-\big(\ii \Delta \hat{\sigma}_z + \ii \Omega(t) \hat{\sigma}_x + \kappa \, \hat{\sigma}_+\hat{\sigma}_-)\frac{\der{t}}{2} + \hat{j}\, \hat{\sigma}_-\right)\ket{\Phi},\\
    &\t{with}\qquad \ket{\Phi} = \int_{-\infty}^{\infty} \sqrt{P_0(j)} \ket{j} \der{j} .
\end{split}
\end{equation}

\subsection{Formal solution of the stochastic dynamics as a filter}
\label{app: filter}

The expression of the infinitesimal time evolution $\hat{\mathds{U}}_{\der t}$ we just derived can be composed for all time intervals to define the evolution over a long period $[0,t]$
\begin{equation}
{\hat{\bf U}}_{t} = \hat{\mathds{U}}_{\der t}(t-\der t) \dots  \hat{\mathds{U}}_{\der t}(\der t) \hat{\mathds{U}}_{\der t}(0),
\end{equation}
where each time step introduces a new quantum system for the photon detection difference measured during the corresponding infinitesimal interval. The joint state of the qubit and all values of the observed homodyne current for $s\in [0,t]$ at the final time $t$ reads
\begin{equation}\begin{split}
    {\hat{\bf U}}_{t} \ket{\psi_0}= \mathcal{T}\left[\exp\left(\int_{0}^t (-\ii \frac{\Delta}{2}\hat{\sigma}_z -\ii \frac{\Omega(s)}{2}\hat{\sigma}_x -\frac{\kappa}{2} \hat{\sigma}_+\hat{\sigma}_- + J(s)\hat{\sigma}_-)\der s\right) \right]\bigotimes_s \ket{\Phi_s}\ket{\psi_0}.
    \end{split}
\end{equation}
Hence, for any fixed values of the measured current $J(s) =j_s/\der s$  we can find the (unnormalized) state of the qubit conditioned on these outcomes 
\begin{equation}
\begin{split}
        &\bigotimes_s \bra{\t{j}_s} {\hat{\bf U}}_{t} \ket{\psi_0}
        =  c_{\bf J} \hat{D}_t \ket{\psi_0} , \quad \t{where} \\
       &\hat{D}_t= \mathcal{T}\left[\exp\left(\int_{0}^t (-\ii \frac{\Delta}{2}\hat{\sigma}_z -\ii \frac{\Omega(s)}{2}\hat{\sigma}_x -\frac{\kappa}{2} \hat{\sigma}_+\hat{\sigma}_- + J(s)\hat{\sigma}_-)\der s\right) \right]
\end{split}
\end{equation}
and $c_{\bf J}=\prod_s \scal{j_s}{\Phi_s}$ is a scalar independent of the input state $\ket{\psi_0}$. Thus, the state of the qubit at time $t$ conditioned on the homodyne detection record reads
\begin{equation}
\ket{\psi_t} \propto \hat{D}_t \ket{\psi_0},
\end{equation} 
and for mixed initial states $\hat{\rho}_t \propto \hat{D}_t \hat{\rho}_0  \hat{D}_t^\dag$. The expression of the operator 
\begin{equation}\label{eq: filter app}
    \hat{D}_t = \mathcal{T}\left[ \exp{
\frac{1}{2} \int_{0}^t 
\begin{pmatrix} -\ii \Delta & -\ii \Omega(s)+2 J(s)\\
-\ii \Omega(s) & \ii \Delta -\kappa
\end{pmatrix} \der s}
    \right]
\end{equation}
can be thought of as a filter relating the record of the drive fields $\bm \Omega_t$ and the values of the measured homodyne current $\bm J_t$ to the map between the states of the qubit at times $0$ and $t$, represented here by a two-by-two complex matrix. In a real experiment $\hat{D}_t $ can be computed by, e.g. discretizing the integral.

Notably, even if the initial state of the qubit is unknown, e.g. $\hat{\rho}_0=\frac{1}{2}\hat{\id}$, after a certain characteristic time the state 
\begin{equation}
    \hat{\rho}_t = \frac{\hat{D}_t \hat{\rho}_0 \hat{D}_t^\dag}{\tr \left[ \hat{D}_t \hat{\rho}_0  \hat{D}_t^\dag \right]}
\end{equation}
becomes pure. 
This is because the stochastic dynamics essentially decouples the state of the system at time $t$ from its state in a far-away past. On the other hand the measured homodyne current $\bm J_t$ reveals information about the unknown initial state and we have just shown how to filter this information, as  
\begin{equation}
    \frac{P(\bm J_t|_{\ket{\psi_0}})}{P(\bm J_t|_{\ket{\phi_0}})} = \frac{\|\hat{D}_t \ket{\psi_0}\|^2}{\|\hat{D}_t \ket{\phi_0}\|^2}
\end{equation}
for any two initial states $\ket{\psi_0}$ and $\ket{\phi_0}$.


\subsection{Derivation of the norm-preserving stochastic Schr\"odinger equation}
\label{App:HD:DerivationNormPreservingSDE}
We start with the stochastic Schr\"odinger equation~\eqref{Eq:Homodyne_NoNorm} which does not preserve the norm of the state $\ket{\psi(t)}$.
Using 
\begin{align}
    \der{\tilde{\rho}} =  |\der{\tilde{\psi}}\rangle\!\tilde{\bra{\psi}} + \tilde{\ket{\psi}}\!\langle\der{\tilde{\psi}}| + |\der{\tilde{\psi}}\rangle\!\langle\der{\tilde{\psi}}| \,,
    \label{Eq:SSEtoSME:Rule}
\end{align}
we can derive the corresponding stochastic quantum master equation
\begin{align}
    \der{\tilde{\rho}} = - \ii  [\hat{H},\tilde{\rho}] \der{t} + \kappa \mathcal{D}[\hat{\sigma}_-] \tilde{\rho} \der{t} + \sqrt{\kappa} J(t) \left( \hat{\sigma}_- \tilde{\rho} + \tilde{\rho} \hat{\sigma}_-^\dagger \right) \der{t} 
    \label{Eq:SME:Unnormalized}
\end{align}
which does not preserve the norm of the density matrix.
Note that the last term in Eq.~\eqref{Eq:SSEtoSME:Rule} contains a contribution of order $\der{t}$ since $\der{W}^2 = \der{t}$.
The last term in Eq.~\eqref{Eq:SME:Unnormalized} shows that the state $\tilde{\rho}$ depends on the homodyne signal $J(t)$, but it does not preserve the norm of $\tilde{\rho}$ during the time evolution. 
This can be compensated by adding a correction term
\begin{align}
    - \sqrt{\kappa} \left( J(t) \der{t} - \der{W} \right) \left( \hat{\sigma}_- \tilde{\rho} + \tilde{\rho} \hat{\sigma}_-^\dagger \right) - \sqrt{\kappa} \ave{\hat{\sigma}_- + \hat{\sigma}_+} \tilde{\rho} \der{W} \,,
\end{align}
which cancels the last term in Eq.~\eqref{Eq:SME:Unnormalized} without introducing any new norm-nonconserving terms.
On the level of the stochastic Schr\"odinger equation, this term can be generated by adding a contribution
\begin{align}
    \left[ - \frac{\kappa}{2} \ave{\hat{\sigma}_- + \hat{\sigma}_+} \hat{\sigma}_- \der{t} - \frac{\kappa}{8} \ave{\hat{\sigma}_- + \hat{\sigma}_+}^2 \der{t} - \frac{\sqrt{\kappa}}{2} \ave{\hat{\sigma}_- + \hat{\sigma}_+} \der{W} \right] \ket{\psi} \,,
\end{align}
which gives rise to Eq.~\eqref{Eq:SDE_homodyne} of the main text.

\section{Neural network architectures and hyperparameters}
\label{App:HP}

\begin{figure}[h!]
	\centering
	\includegraphics[width=0.9\linewidth,angle=0]{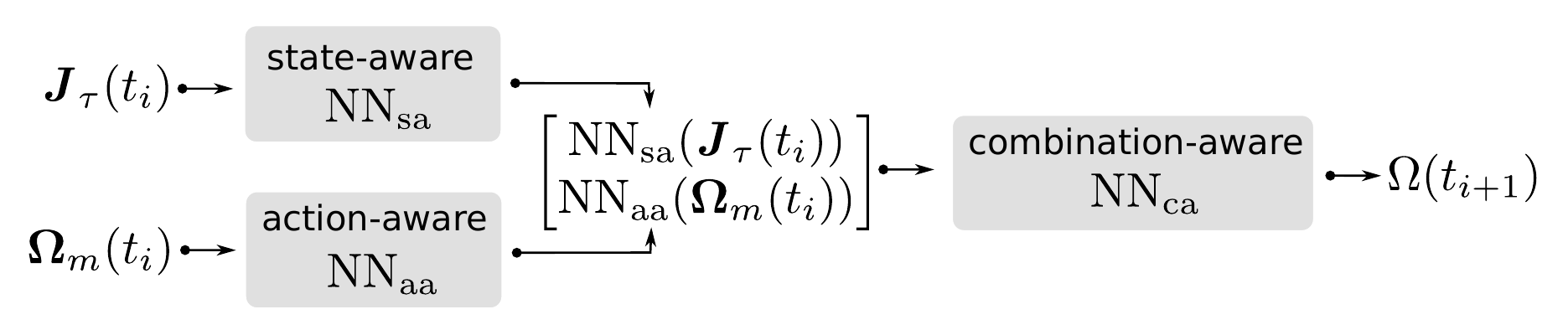}
	\caption{Scheme of the NN used in Section~\ref{Sec:SDE_control_Jhom} which consists of three segments of fully connected networks.
	The first one, the state-aware NN$_{\rm sa}$, obtains the vector $\bm{J}_\tau(t_i)$ which bears the information on the state of the controlled system in form of its quadrature $\ave{\sigma_x}$ over the time interval $[t_i-\tau, t_i]$.
	Similarly, the action-aware NN$_{\rm aa}$ takes as an input a vector $\bm{\Omega}_m(t_i)$ of the last $m$ actions taken prior to the time $t_i$.
	The output of NN$_{\rm sa}$ and NN$_{\rm aa}$ is concatenated into a vector and enters the last segment, which we refer to as combination-aware NN$_{\rm ca}$.
	The final output is the next drive value $\Omega(t_{i+1})$ to be applied.
    The parameters of the network can be found in Tab.~\ref{Tab:HP}.
	}
	\label{Fig:Jhom_NN}
\end{figure}

The hyperparameters used in the control tasks are summarized in Tab.~\ref{Tab:HP}.
The architecture of the NN used in Section~\ref{Sec:SDE_control_Jhom} is shown in Fig.~\ref{Fig:Jhom_NN}.
In all NNs, we use ReLUs as activation functions for hidden layers and the softsign activation function for the last layer.
Modifications of this simple architecture, e.g. through the application of recurrent neural networks to capture temporal correlations in the homodyne current for the SDE control in Section~\ref{Sec:SDE_control_Jhom}, might be essential for the control of complex many-body quantum systems.

In this work, we used NNs as universal function approximators.
While this approach is very general, other choices of a controller, e.g. Chebychev polynomials or Fourier basis expansions, could boost the performance for low-dimensional inputs as in the case of a qubit, because they can be optimized to the problem at hand.
The use of the sparse identification for dynamical system method~\cite{Brunton2016, boninsegna2018, kaiser2018sparse} or symbolic regression tools~\cite{cranmer2020discovering} could further allow one to replace the trained NNs by a symbolic description based on a pre-defined library of operators.

\begin{table*}
	\centering
		\begin{tabular}{@{}rrrcrcr@{}}\toprule
			& \multicolumn{2}{c}{$\ket{\psi(t)}$ (SDE)} & \phantom{abc}& \multicolumn{1}{c}{$J(t)$ (SDE)} & \phantom{abc} & \multicolumn{1}{c}{$\ket{\psi(t)}$ (ODE) }\\
			\cmidrule{2-3} \cmidrule{5-5} \cmidrule{7-7} 
			& Section~\ref{Subsec:Continuous} & Section~\ref{Subsec:Piecewise} && Section~\ref{Sec:SDE_control_Jhom} && \ref{App:ODE}\\\midrule
			\midrule
			\multicolumn{5}{l}{solver}\\
			scheme & EulerHeun & RKMil && RKMil && Tsit5 \\
			$N_{\rm sub}$ & 200 & 20 && 80 && \phantom{200} \\
			$\der t$ & $10^{-4}  \kappa^{-1}$ & $10^{-3}  \kappa^{-1}$ && $2.5 \cdot10^{-4}  \kappa^{-1}$ && adaptive \\
			\midrule
			\multicolumn{5}{l}{loss function}\\
			$c_F$ & 1 & 0.8  &&  1.2 && 1 \\
			$c_{F50}$ & 0 &  $1.8 N/50$ && $0.8 N/50$ && 0 \\
			$c_{\Omega}$& 0 & $10^{-3}$ && $10^{-3}$ && 0  \\
			\midrule
			\multicolumn{5}{l}{Adam optimizer} \\
			learning rate & 0.0015 & 0.0001 &&  0.0001 && 0.0015 \\
			batchsize $b$& 64 & 64 && 64 &&  256\\
			epochs & 1000 & 3000 && 14000 && 400 \\
			\midrule
			\multicolumn{5}{l}{NN  }\\
			LLS 1 &(4, 256)  & (4, 256)  && ($N_{\rm sub}$, 256) && (4, 256) \\
			LLS 2 &(256, 64) & (256, 128) && (256, 256) && (256, 64) \\
			LLS 3 &(64, 1) &  (128,64) && (256, 128)  && (64, 1) \\
			LLS 4 &\phantom{(4, 256)}&(64, 1)  && \phantom{(4, 256)} && \phantom{(4, 256)}\\
			LLA 1 &\phantom{(4, 256)}&\phantom{(4, 256)}  &&  ($N_{\rm sub}$/10, 128) && \phantom{(4, 256)}\\
			LLA 2 &\phantom{(4, 256)}&\phantom{(4, 256)}  &&  (128, 128) && \phantom{(4, 256)}\\
			LLC 1 &\phantom{(4, 256)}&\phantom{(4, 256)}  && (256, 64) && \phantom{(4, 256)}\\
			LLC 2 &\phantom{(4, 256)}&\phantom{(4, 256)}  && (64, 32) && \phantom{(4, 256)}\\
			LLC 3 &\phantom{(4, 256)}&\phantom{(4, 256)}  && (32, 1) && \phantom{(4, 256)}\\		
			\bottomrule
	\end{tabular}
	\caption{Hyperparameters employed to train the neural networks in the case of continuously updated control parameters (Section~\ref{Subsec:Continuous}), piecewise-constant control parameters with knowledge of the state $\ket{\psi}$ (Section~\ref{Subsec:Piecewise}),  piecewise-constant control parameters with knowledge of the homodyne current (Section~\ref{Sec:SDE_control_Jhom}), and the closed-system dynamics described by the Schr\"odinger equation (\ref{App:ODE}).
	The number of checkpoints is $N=150$ for all simulations.
	The value of $N_{\rm sub}$ gives the number the solver substeps between the checkpoints.
	The coefficient $c_{F50}$ refers to the modification of the loss function~\eqref{Eq:Infidelity} which is limited to the last $50$ steps of the control interval.
	Specifications of the solvers are provided in the SciML documentation~\cite{rackauckas2017differentialequations}.
	\lq LLS~$n$\rq, \lq LLA~$n$\rq\, and \lq LLC~$n$\rq\ denotes the $n$th  linear-layer state-aware, action-aware, and combination-aware network, respectively.
	The numbers of input and output channels of the layers are specified in the brackets.}
	\label{Tab:HP}
\end{table*}

\section{Continuous adjoint sensitivity method for ODEs and SDEs}
\label{App:ODE_and_Implementation}

In this appendix, we discuss the continuous adjoint sensitivity method for ordinary differential equations (ODEs) and its generalization to SDEs used in the main text. 
In~\ref{App:ODE}, the continuous adjoint sensitivity method for ODEs is used to compare the stochastic control scenario discussed in Section~\ref{Subsec:Continuous} with the associated unitary control scenario in the case of a closed system.
In~\ref{App:Implementation}, we provide an intuitive understanding for the stochastic adjoint process and discuss technical details regarding the continuous adjoint sensitivity method for SDEs and its implementation~\cite{GSOC}. 
In all implementations, we use an isomorphism to map the complex amplitudes of the quantum state to real numbers, as required by the AD backend~\cite{innes19}. 

\subsection{Quantum control of a qubit in a closed system}
\label{App:ODE}

In this section, we aim at controlling the closed-system dynamics of a qubit given by the Schr\"odinger equation 
\begin{equation}
 \der{\ket{\psi(t)}}= - \ii H^{\rm CS} \ket{\psi(t)}  \der{t} =: K^{\rm CS}_{\psi(t)} \der{t}\,,
\label{Eq:App:Schroedinger}
\end{equation}
with Hamiltonian
\begin{equation}
 H^{\rm CS}=\frac{\Delta}{2} \sigma_z +\frac{\Omega(t)}{2} \sigma_x \,,
\label{Eq:App:H}
\end{equation}
where $\Delta$ is the qubit transition frequency~\cite{schaefer20}. 
As in Section~\ref{Subsec:Continuous}, we choose a NN with parameters $\vec{\theta}$ as the controller ansatz and we allow the NN to change the control drive $\Omega(t)$ in every time step based on the state $\ket{\psi(t)}$. 
The initial states are uniformly distributed on the Bloch sphere, see Eq.~\eqref{Eq:initial_state}.
We compute the forward pass, Eq.~\eqref{Eq:App:Schroedinger}, with the adaptive Tsitouras 5/4 embedded Runge-Kutta (Tsit5) scheme as implemented in the DifferentialEquations.jl package~\cite{rackauckas2017differentialequations}. 
Given this solution, we use the loss function~\eqref{Eq:Loss} with weights $c_F = 1, c_\Omega = 0$. 
As discussed in Section~\ref{Subsec:Differentiable}, it depends on the checkpoints  at times $\{t_i\}_{i=1}^N$, $\mathfrak{L} = \mathfrak{L}(\{\ket{\psi(t_{i})}\})$.

As discussed in Section~\ref{Subsec:Adjoint}, the continuous adjoint sensitivity method circumvents the memory issues of discrete reverse-mode AD and scales better with the number of parameters than forward-mode AD.
To derive this adjoint method, one first adds a zero to the loss function, Eq.~\eqref{Eq:Loss}, and rewrites it as a time integral
\begin{equation}
   I(\vec{\theta})=\int_{t_0}^{t_N}\left[ \frac{1}{N} \sum_{i=0}^N \left(1-\abs{\scal{\psi(t_i)}{\psi_{\rm tar}}}^2\right)\delta(t-t_i) - \vec{\lambda}^{\dagger}(t)\left(\dot{\ket{\psi}} - K^{\rm CS}_{\psi(t)}(\vec{\theta}) \right) \right] \der{t}\,,
\label{Eq:App:loss1}    
\end{equation}
where we inserted $\mathfrak{L}_F$ defined in Eq.~\eqref{Eq:Infidelity} and introduced the Lagrange multiplier $\vec{\lambda}$, such that
$I(\vec{\theta})=\mathfrak{L}(\vec{\theta})$ and $\nabla_{\vec{\theta}} I(\vec{\theta})=\nabla_{\vec{\theta}}\mathfrak{L}(\vec{\theta})$.
After an integration by parts and re-arrangement of terms for $\nabla_{\vec{\theta}} I(\vec{\theta})$, one finds that computing the gradients $\nabla_{\vec{\theta}}\mathfrak{L}$ requires to evaluate the time evolution of the quantity $\vec{a}_{\psi}(t) = \vec{\lambda}^\dagger(t)$.
This leads to the gradients $\nabla_{\psi(t)} \mathfrak{L}$ of the loss function with respect to the state $\ket{\psi(t)}$ for all times $t$ and is called the adjoint process: 
\begin{equation}
     \vec{a}_{\psi}(t)=\nabla_{\psi(t)} \mathfrak{L} \,.
\end{equation}
The associated adjoint ODE problem satisfies the differential equation~\cite{rackauckas2020universal,pontryagin2018, chen2018neural, johnson2007, jia2019neural} 
\begin{align}
    \der{\vec{a}_{\psi}}(t)&= -\left(\vec{a}_{\psi}^\dagger(t) \cdot     \nabla_{\psi(t)}\right) K^{\rm CS}_{\psi(t)}\der{t}\,,\nonumber\\ 
    \vec{a}_{\psi}(t_{0})&= \vec{a}_{\psi}(t_{N}) + \int_{t_{N}}^{t_{0}} \left[\frac{d \vec{a}_{\psi}(t)}{dt} - \sum_{i \neq N} \delta(t-t_{i}) \nabla_{\psi(t_i)} \mathfrak{L}\right] \der{t} \,,
  \label{Eq:App:adjoint_ODE}
\end{align}
with the initial condition 
\begin{align}
    \vec{a}_{\psi}(t_N)=\nabla_{\psi(t_N)} \mathfrak{L}(\{\ket{\psi(t_i)}\}) \,.
\end{align}

This adjoint ODE is defined backwards in time from $t_N$ to $t_0$.
To compute the vector-Jacobian products in Eq.~\eqref{Eq:App:adjoint_ODE}, one needs to know the value of the state $\ket{\psi(t)}$ along its entire trajectory, which has been computed in the forward pass.
Thus, we must store these states or recompute them by solving the ODE backwards in time starting from the final value $\ket{\psi(t_N)}$,
\begin{equation}
    \ket{\psi(t)} =  \ket{\psi(t_N)} + \int_{t_N}^{t} K^{\rm CS}_{\psi(t')} \der{t'}\,
    \label{Eq:App:backwardODE}
\end{equation}
together with the adjoint process.
This does not introduce a significant memory overhead.
Computing the gradients $\nabla_{\vec{\theta}}\mathfrak{L}$ requires yet another integral, which depends on the original and the adjoint process, $\ket{\psi(t)}$ and $\vec{a}_{\psi}(t)$, respectively.
With the initial condition
\begin{align}
    \vec{a}_{\vec{\theta}}(t_N)= \vec{0}_{{\rm Dim}[\theta]} \,,
\end{align}
the gradients $\nabla_{\vec{\theta}}\mathfrak{L}=\vec{a}_{\vec{\theta}}(t_0)$ are determined by
\begin{align}
    \der{\vec{a}}_{\vec{\theta}}(t)&= -\left(\vec{a}_{\psi}^\dagger(t) \cdot     \nabla_{\vec{\theta}}\right) K^{\rm CS}_{\psi(t)}\der{t}\,,\nonumber\\ 
    \vec{a}_{\vec{\theta}}(t_{0})&=  \vec{a}_{\vec{\theta}}(t_N) + \int_{t_{N}}^{t_{0}} \frac{d \vec{a}_{\vec{\theta}}(t)}{dt}  \der{t} \,,
\label{Eq:App:adjoint_ODE2}
\end{align}

Therefore, the gradients of the loss function $\nabla_{\vec{\theta}}\mathfrak{L}$ with respect to the neural network parameters $\vec{\theta}$ can be obtained by a single (adjoint) ODE with an augmented state given by $\ket{\psi(t)}$, $\vec{a}_{\psi}(t)$, and $\vec{a}_{\vec{\theta}}(t)$. 

Because of the reversion of the ODE, Eq.~\eqref{Eq:App:backwardODE} may be numerically unstable.
Therefore, we use an interpolating adjoint algorithm to increase stability~\cite{rackauckas2020universal}. 
When solving the augmented state backwards in time, we recompute the forward pass sequentially for all time intervals $[t_i, t_{i+1}]$ between two checkpoints.
Then, fourth-order interpolations of the recomputed forward pass are used to compute the vector-Jacobian products for the reverse pass. 

\begin{figure}[h!]
	\centering
	\includegraphics[width=1.0\linewidth, angle=0]{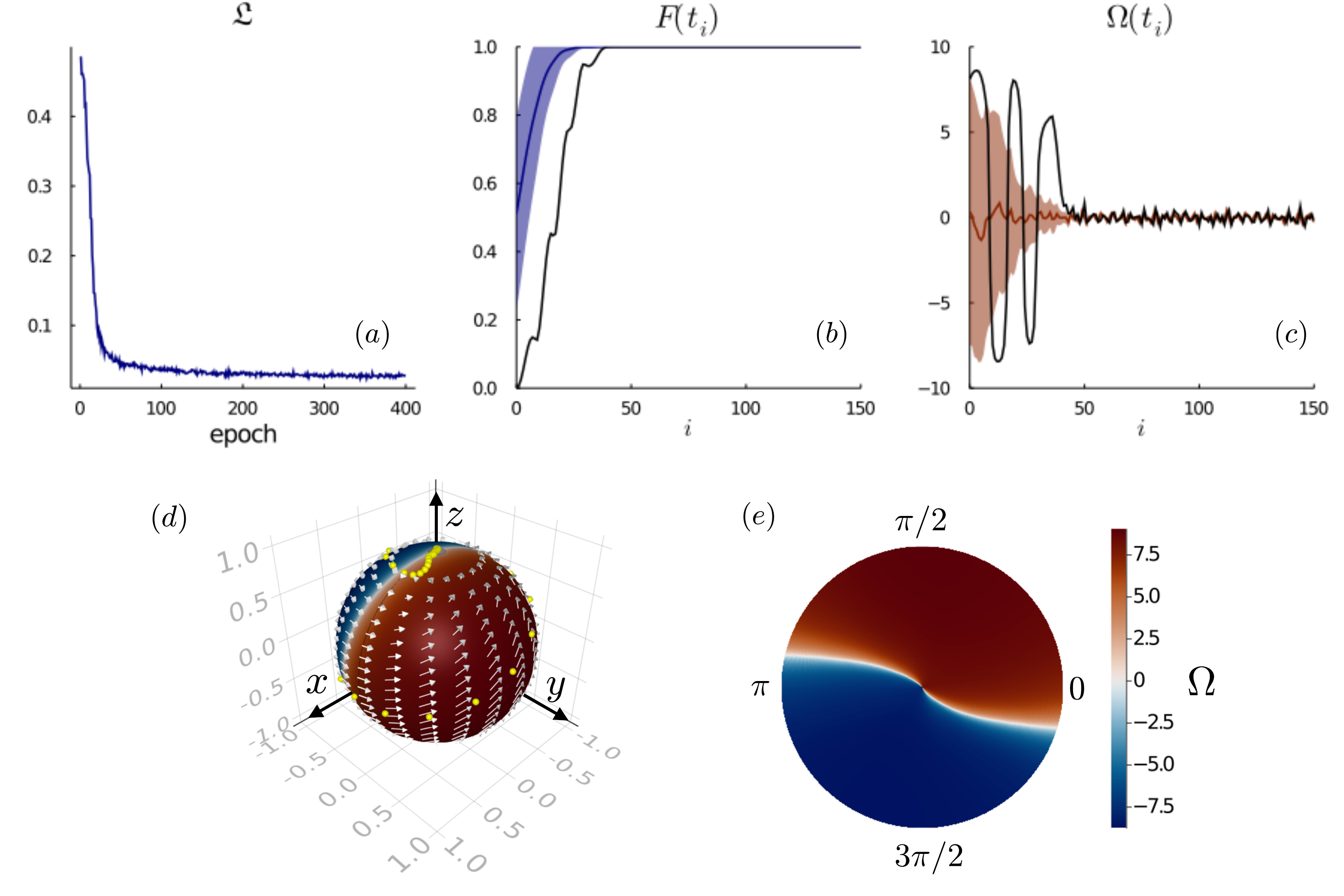}
	\caption{
	Preparation of the state $\ket{e}$ in a closed system described by the Hamiltonian, Eq.~\eqref{Eq:App:H}, based on RL and \DP{}.
	(a) Loss function, Eq.~\eqref{Eq:Loss}, as a function of training epochs.
	Means (blue/red) and standard deviations (shaded) for (b) fidelity and (c) control parameters, respectively, as a function of the steps $i$ at which checkpoints are stored.
    The black line in (c,d) as well as the yellow dots in (d) highlight the special case of the initial state $\ket{g}$.
    (d) Bloch sphere with color coding representing $\Omega$ and vector field showing the infinitesimal evolution of the state $\ket{\psi}$ at that position.
    (e) Stereographic projection of the southern hemisphere of the Bloch sphere according to Eq.~\eqref{Eq:stereographic_projection}.
    The employed hyperparameters can be found in Tab.~\ref{Tab:HP} in \ref{App:HP}.}
	\label{Fig:AppendixB}
\end{figure}

The results of the closed-system control task are shown in Fig.~\ref{Fig:AppendixB}. 
We observe a very fast and smooth learning process with our physics-informed RL framework.
The target state $\ket{e}$ is an eigenstate of the Hamiltonian~\eqref{Eq:App:H} for $\Omega = 0$, therefore, the control drive $\Omega$ is switched off once the target state $\ket{e}$ is reached.
Figures~\ref{Fig:AppendixB}(d) and (e) show that the control strategy, illustrated on the Bloch sphere and with a stereographic projection, resembles the stochastic case discussed in the main text.

\subsection{Technical details of the continuous adjoint sensitivity method for SDEs}
\label{App:Implementation}

To generalize the continuous adjoint sensitivity method for ODEs to It\^o SDEs, one first needs to figure out how the sample path of an SDE can be reversed, i.e. how one can reconstruct the forward pass of the state $\ket{\psi(t)}$ from time $t_0$ to $t_N$ by a reversed time evolution from $t_N$ to $t_0$ launched at $\ket{\psi(t_N)}$.
The reversion of an SDE 
\begin{equation}
   \s_der{\ket{\psi}}= \tilde{K}_{\psi(t)}\der{t}+ M_{\psi(t)}\circ \der{W}(t) \,, 
    \label{Eq:App:SDEStrat}
\end{equation}
defined in the Stratonovich sense, is given by~\cite{li2020, kidger2021}
\begin{equation}
    \ket{\psi(t)} =  \ket{\psi(t_N)} + \int_{t_N}^{t} \tilde{K}_{\psi(t')} \der{t'} + \int_{t_N}^{t} M_{\psi(t')} \circ \der{W(t')} \,,
    \label{Eq:App:backwardSDE}
\end{equation}
with noise values $W(t)$ identical to those sampled in the forward pass.
This closely resembles the reversion in case of an ODE shown in Eq.~\eqref{Eq:App:backwardODE}. 
To restore the noise, we use a dense noise grid of the noise values used in the forward pass.
The memory overhead caused by using a noise grid could be traded against speed by using a virtual Brownian tree~\cite{li2020} or a Brownian interval~\cite{kidger2021}, which enable the reconstruction of $W(t)$ by storing only very little information such as the seed of the employed pseudo-random number generator.
Despite the allocation of the noise values, the continuous stochastic adjoint sensitivity method is still much more memory efficient than a discrete reverse-mode AD backpropagation through the solver operations. 

From the reverse Stratonovich SDE, Eq.~\eqref{Eq:App:backwardSDE}, we can straightforwardly obtain the reverse It\^o SDE 
\begin{equation}
    \ket{\psi(t)} =  \ket{\psi(t_N)} + \int_{t_N}^{t} \left(K_{\psi(t')} -2C^{\rm IS}_{\psi(t)} \right)\der{t'} + \int_{t_N}^{t} M_{\psi(t')} \der{W(t')} \,
    \label{Eq:App:backwardSDEIto}
\end{equation}
of the monitored qubit, Eq.~\eqref{Eq:SDE_homodyne}, where the standard conversion rule
\begin{equation}
      C^{\rm IS}_{\psi(t)}=\frac{1}{2} \left(M_{\psi(t)}\cdot \nabla_{\psi(t)}\right) M_{\psi(t)}
\end{equation}
in Eq.~\eqref{Eq:App:backwardSDEIto} accounts for the required transformation from It\^o to the Stratonovich sense and vice versa~\cite{kloeden2013numerical}. 
  
Analogously to the ODE case, taking the scalar loss function $\mathfrak{L}$, Eq.~\eqref{Eq:Loss}, the adjoint process
\begin{equation}
  \vec{a}_{\psi}(t)=\nabla_{\psi(t)} \mathfrak{L}(\{\ket{\psi(t_i)}\}) 
\end{equation}
to compute the gradients of $\mathfrak{L}$ with respect to the state $\ket{\psi(t)}$ is now a strong solution~\cite{kloeden2013numerical} of the adjoint It\^o SDE 
\begin{align}
   \der{\vec{a}_{\psi}}(t)=& -\left(\vec{a}_{\psi}^\dagger(t) \cdot \nabla_{\psi(t)}\right) \left(K_{\psi(t)} - 2C^{\rm IS}_{\psi(t)}\right) \der{t}-\left(\vec{a}_{\psi}^\dagger(t) \cdot \nabla_{\psi(t)}\right) M_{\psi(t)}  \der{W(t)}
\label{Eq:App:adjoint_SDE}
\end{align}
with the initial condition 
\begin{align}
   \vec{a}_{\psi}(t_N) = \nabla_{\psi(t_N)} \mathfrak{L}(\{\ket{\psi(t_i)}\}) \,.
\end{align}
Again, the value of the state $\ket{\psi(t)}$ of the forward pass is seen to be embedded in the vector-Jacobian products within the backward pass and, therefore, the knowledge of the state $\ket{\psi(t)}$ along its trajectory is necessary. 
Using Equation~\eqref{Eq:App:backwardSDEIto}, we can recompute the state $\ket{\psi(t)}$ without having to store the full trajectory of the forward pass.

This SDE can be augmented by the additional quantity $\vec{a}_{\vec{\theta}}(t)$ to compute the gradients $\vec{a}_{\vec{\theta}}(t_0) = \nabla_{\vec{\theta}}\mathfrak{L}$.
It is the solution of the stochastic differential equation
\begin{align}
     \der{\vec{a}_{\vec{\theta}}}(t)=& -\left(\vec{a}_{\psi}^\dagger(t) \cdot \nabla_{\vec{\theta}}\right) \left(K_{\psi(t)} - 2C^{\rm IS}_{\psi(t)}\right) \der{t}-\left(\vec{a}_{\psi}^\dagger(t) \cdot \nabla_{\vec{\theta}}\right) M_{\psi(t)}  \der{W(t)}
\label{Eq:App:adjoint_SDE2}
\end{align}
with the initial condition
\begin{equation}
    \vec{a}_{\vec{\theta}}(t_N)= \vec{0}_{{\rm Dim}[\theta]} \,.
\end{equation}

As a consequence of the scalar noise character of the forward SDE, Eq.~\eqref{Eq:SDE_homodyne}, the adjoint SDE with an augmented state according to Eqs.~\eqref{Eq:App:backwardSDEIto}, \eqref{Eq:App:adjoint_SDE}, and~\eqref{Eq:App:adjoint_SDE2} also has scalar noise. 
Similar to the ODE setting, solving SDEs backwards is not guaranteed to be stable. 
Thus, to improve the stability, we modify this approach by resetting the reverse integration using the checkpoints $\{\ket{\psi(t_i}\}$.

\section{The effect of the decay rate $\kappa$ on the fidelity $F$}
\label{App:kappa}

\begin{figure}[h!]
	\centering
	\includegraphics[width=1.0\linewidth, angle=0]{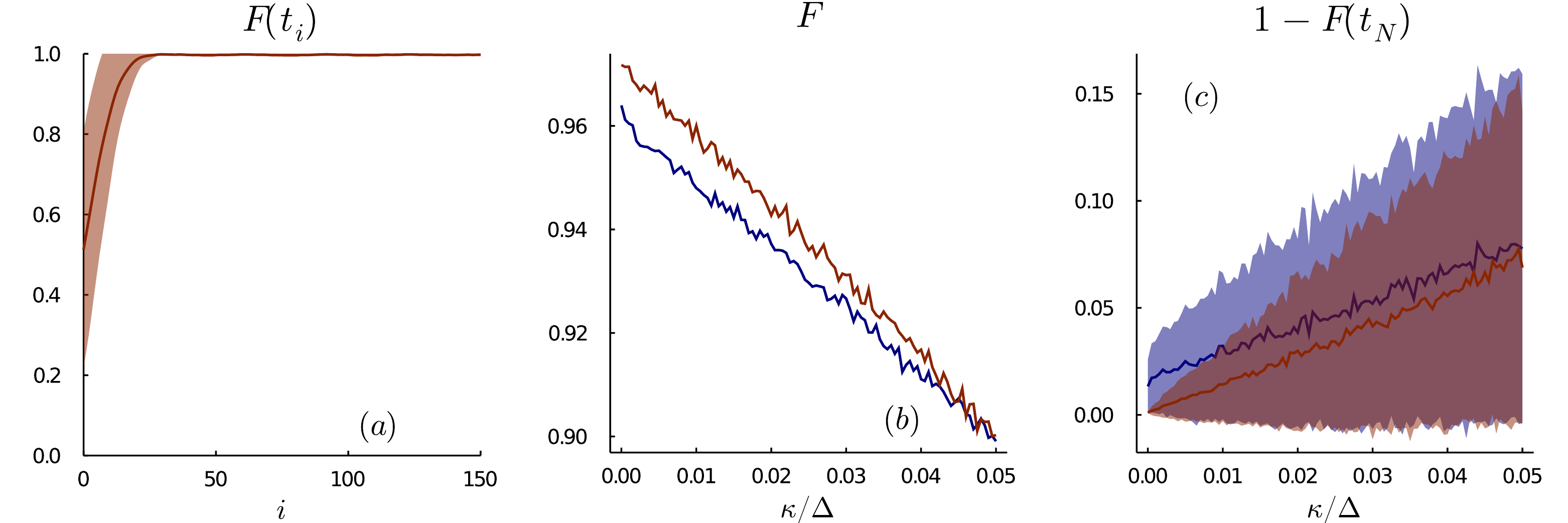}
	\caption{
	Comparison of the preparation of the state $\ket{e}$ for the continuously monitored qubit described by the SDE~\eqref{Eq:SDE_homodyne}, based on continuously updated control parameters for different values of $\kappa/\Delta$.	
	(a) The solid red line shows the mean fidelity $F(t_i)$ as a function of the time steps $t_i$ at which checkpoints are stored in the case of $\kappa/\Delta=0.001$ and the neural network trained using the continuous adjoint sensitivity method with $\kappa/\Delta=0.05$ of Sec.~\ref{Subsec:Continuous} (red).
	The shaded region represents the corresponding standard deviation.
	(b) Average fidelity over all time steps $i$ as a function of $\kappa/\Delta$ for the neural network from (a) (red) and the hand-crafted control strategy of Sec.~\ref{Subsec:Hand-crafted} (blue).
	(c) Average final-state infidelity (solid lines) and corresponding standard deviation (shaded regions) as a function of $\kappa/\Delta$ for the neural network from (a) (red) and the hand-crafted control strategy of Sec.~\ref{Subsec:Hand-crafted} (blue).
	The mean values and standard deviations are computed for a set of 512 randomly sampled initial states on the Bloch sphere. 
    The hyperparameters are listed in Tab.~\ref{Tab:HP}.}
	\label{Fig:AppendixD}
\end{figure}

In the main text, we report mean fidelities of about $F\approx0.9$ averaged over the whole control interval including the very low fidelities due to the transient initial dynamics.
This value of the fidelity is determined by the physical parameters of the quantum system ($\Delta$, $\kappa$) as well as by the experimental limitations of the control scheme (e.g. $\Omega_{\rm max}$ and the feedback control frequency).
We note that these parameters are setup-specific and do not represent a hard fidelity limit for our proposed control scheme.
In Figure~\ref{Fig:AppendixD}, we demonstrate that the mean fidelity over the full time interval as well as the final state infidelity can easily be improved by decreasing the decay rate $\kappa$ with respect to the detuning $\Delta$.
In the limit $\kappa/\Delta \to 0$, one recovers a fidelity close to unity comparable to the closed-system case, see~\ref{App:ODE}.
We further see that the difference between the trained NN and the hand-crafted strategy gets more pronounced as $\kappa$ is decreased.

\section*{References}
\bibliography{References}
\end{document}